\documentclass {article}
\usepackage{graphicx}

\begin{document}

\title{Holostar Thermodynamics}

\author{Michael Petri\thanks{email: mpetri@bfs.de} \\Bundesamt f\"{u}r Strahlenschutz (BfS), Salzgitter, Germany}

\date{{June 16, 2003 (v1)} \newline {May 1, 2004 (v3)}}

\maketitle

\begin{abstract}

A simple thermodynamic model for the final state of a collapsed,
spherically symmetric star is presented. It is assumed that at the
end-point of spherically symmetric collapse the particles within
the star become ultra relativistic and that their thermodynamic
properties can be described by an ideal gas of ultra-relativistic
fermions and bosons. The metric at the final stage of collapse is
assumed to approach the static metric of the so called holographic
solution, a new exact spherically symmetric solution to the
Einstein field equations with zero cosmological constant.

If the geometry of a collapsed star can be described by the
holostar metric, the established picture of gravitational collapse
of a relativistic star changes significantly. The metric-induced
"expansion" of space in combination with the quantum mechanical
degeneracy pressure of its constituent matter allow a collapsing
star to settle down to a thermodynamically stable state, which
doesn't have a point-mass at its center, irrespective of the
star's mass. The thermodynamical configuration of the holostar is
stabilized by a non-zero chemical potential of the relativistic
fermions, which is proportional to the local radiation temperature
within the holostar. The non-zero chemical potential acts as a
natural source for a significant matter-antimatter asymmetry in a
self gravitating object.

The final configuration has a radius slightly exceeding the
gravitational radius of the star. The radial coordinate difference
between gravitational and actual radius is of order of the Planck
length and roughly proportional to the square root of the
effective number of degrees of freedom of the particles within.
The total number of ultra-relativistic particles within the star
is proportional to its proper surface-area, measured in units of
the Planck-area. The entropy per particle $\sigma$ can be calculated
from the thermodynamic model. $\sigma >\simeq \pi$, nearly independent
from the specifics of the thermodynamic model. This is first
direct evidence for the microscopic-statistical nature of the
Hawking entropy and indicates, that the holographic principle is
valid for compact self gravitating objects of any size.

The free energy within the holostar is minimized to $F = 0$,
meaning that - on average - the entropy per particle is equal to
the energy per particle per temperature, i.e. $\sigma = \epsilon / T$.

A "Stefan-Boltzmann-type" relation between the local surface
temperature and the proper surface area of the star is derived.
This relation implies a well defined internal radiation
temperature proportional to $1/ \sqrt{r}$. The red-shift factor of
the holostar's surface with respect to an observer at infinity is
proportional to $1 / \sqrt{M}$, so that the holostar's temperature
at infinity is proportional to $1/M$. The Hawking temperature and
-entropy laws are derived from microscopic statistical
thermodynamics up to a constant factor.

The factor relating the holostar's temperature at infinity to the
Hawking temperature can be expressed in terms of the internal
local radiation temperature and the local total matter-density,
allowing an experimental determination of this factor. Assuming,
that the universe can be described by a large holostar and using
the recent experimental data for the CMBR-temperature and the
total matter-density the Hawking formula is verified to an
accuracy better than 1\%.

The surface area of a holostar consists of a two dimensional
membrane with high tangential pressure. The membrane's
contribution to the entropy and gravitating mass of the holostar
is discussed briefly under the assumption, that the membrane
consists of a gas of bosons whose number is roughly equal to the
number of the interior particles.

A table for the relevant thermodynamic parameters for various
combinations of the degrees of freedom and the chemical potentials
of the interior particles (fermions and bosons) is given. The
total number of particle degrees of freedom at ultra-relativistic
energies is estimated to be $f \approx 7000$. The holographic
solution preserves the relative ratios of the energy-densities of
the fundamental particle species. This allows an estimation of the
proton to electron mass-ratio $m_p / m_e \approx f/4$.

The case of a "zero-temperature" holostar and some properties
expected from a rotating holostar are discussed briefly.
\end{abstract}

\section{\label{sec:Intro}Introduction:}

In \cite{petri/bh} a variety of new solutions to the field equations
of general relativity were derived. These solutions describe a
spherically symmetric compact gravitating object with a generally
non-zero interior matter-distribution. The boundary of the matter
distribution generally lies outside of the object's gravitational
radius and consists of a two-dimensional spherical membrane of
high surface-tension/pressure, whose energy-content is comparable
to the gravitating mass of the object.

Some of the new solutions don't contain a point-singularity at the
origin, indicating that spherically symmetric collapse of a large
star or a galactic core might not necessarily end up in a black
hole of the Schwarzschild (vacuum) type, if the pressure becomes
anisotropic.

The new solutions cover a wide range of possibilities. The most
promising solutions appear to be those characterized by a
mass-density of $\rho \propto 1 / (8 \pi r^2)$ within the
interior.

The solution with $\rho = 1/(8 \pi r^2)$, the so called
holographic solution, or short "holostar", is of particular
interest. Its geometric properties have been discussed in
\cite{petri/hol} in some detail. The holostar is characterized by
the property, that the stress-energy content of the holostar's
membrane is equal to its gravitating mass. Alternatively the
holostar's total gravitating mass can be derived by an integral
over the {\em trace} of the stress-energy tensor, a Lorentz
invariant quantity.

The interior matter-state of the holographic solution can be
interpreted as a collection of radially outlayed strings, with the
end-points of the strings attached to the holostar's boundary
membrane. Every string segment attached to the membrane occupies a
membrane segment of exactly one Planck area. The interior strings
are densely packed, their mutual transverse separation is exactly
one Planck area. This dense package might be at the heart of the
explanation, why the holostar doesn't collapse to a singularity,
regardless of it's size. For a more detailed discussion of the
string nature of the holographic solution see \cite{petri/string}.

Although the holographic solution has a strong string character,
its interior matter state can also be interpreted in terms of
particles. In this paper a simple thermodynamic model for the
interior matter state is explored, which provides a genuine
microscopic statistical explanation for the Hawking temperature
and entropy of a compact self gravitating object and gives a
thermodynamic explanation for the matter-antimatter asymmetry in
curved space-time.

\section{\label{sec:holo:intro}A short introduction to the holographic solution}

The holographic solution is an exact solution to the Einstein
field equations with zero cosmological constant\footnote{The
holographic solution can be viewed as the simplest possible
solution to the field equations. Although it contains matter, it
is much simpler than a pure vacuum solution. How can this be?
General relativity is a {\em non-linear} theory. Even its {\em
vacuum} equations are non-linear and all of its known solutions
(=classical black holes) quite complicated. Therefore most of
general relativity's practitioners tend to think that {\em all}
relevant solutions to the field equations {\em must} be non-linear
and highly complex. There is a general feeling that the solutions
will become even more complicated when matter is introduced into
the theory. Linearizing the field equations is viewed as a mere
approximation, appropriate only in the weak field limit. The full
result should always be non-linear. But this is not the case: The
introduction of {\em string type} matter into the general theory
of relativity {\em simplifies} the field equations in a very
essential way. In the spherically symmetric case it is very easy
to see that the field equations are {\em linearized}, if the
matter follows a string equation of state $\rho = - P_r$. That
string type matter {\em must} reduce the complexity of the field
equations becomes clear, if one realizes that the active
gravitational mass-density of a string is always {\em zero}.
Neither the vacua of the classical black hole solutions nor other
types of matter have this property. But the active gravitational
mass density can be viewed as the "true" source of the
gravitational field. The equations for the local proper geodesic
acceleration in a local Minkowski frame have the active
gravitational mass-density as a source term.}. The spherically
symmetric metric of the holographic solution has been derived in
\cite{petri/bh} :

\begin{equation}
ds^2 = g_{tt}(r) dt^2 - g_{rr}(r) dr^2 - r^2 {d\Omega}^2
\end{equation}

\begin{equation} \label{eq:A}
g_{tt}(r) = 1/g_{rr}(r) = \frac{r_0}{r}(1-\theta(r-r_h)) +
(1-\frac{r_+}{r})\theta(r-r_h)
\end{equation}

with

$$ r_h = r_+ + r_0$$
$$ r_+ = 2M$$

All quantities are expressed in geometric units $c=G=1$. For
clarity $\hbar$ will be shown explicitly. $\theta$ and $\delta$
are the Heavyside-step functional and the Dirac-delta functional
respectively. $r_h$ denotes the radial coordinate position of the
holostar's surface, which divides the space-time manifold into an
interior source region with a non-zero matter-distribution and an
exterior vacuum space-time. $r_+$ is the radial coordinate
position of the gravitational radius (Schwarzschild radius) of the
holostar. $r_+$ is directly proportional to the gravitating mass
$M = r_+/2$. $r_0$ is a fundamental length
parameter.\footnote{$r_0$ has been assumed to be roughly twice the
Planck-length in \cite{petri/bh, petri/hol}. The analysis in
\cite{petri/charge} indicates $r_0^2 \simeq 4 \sqrt{3/4}$ at low
energies. In this paper a more definite relationship in terms of
the total number of particle degrees of freedom at high
temperatures will be derived.}

The matter fields (mass density, principal pressures) of any
spherically symmetric gravitationally bound object can be derived
from the metric by simple differentiation (see for example
\cite{petri/bh}). For the discussion in this paper only the {\em
radial} metric coefficient $g_{rr}(r)$ is essential. In the
spherically symmetric case the total mass-energy density $\rho$
can be calculated {\em solely} from the radial metric
coefficient.\footnote{If the principal pressures are to be derived
from the metric, one must also know the {\em time-coefficient} of
the metric. This can be loosely interpreted such, that pressure is
a "time-dependent" phenomena, i.e. the physical origin of pressure
is intimately related to the (unordered) motion of the particles,
whereas the mass-energy density of a static body not necessarily
needs the concept of motion for it's definition.} For any
spherically symmetric self gravitating object the following
general relation holds:

\begin{equation} \label{eq:r/grr}
{\left(\frac{r}{g_{rr}}\right)}' = 1 - 8 \pi r^2 \rho
\end{equation}

It is obvious from the above equation, that a matter-density $\rho
= 1 / (8 \pi r^2)$ is special. It renders the differential
equation for $g_{rr}$ homogeneous and leads to a strictly linear
dependence between $g_{rr}$ and the radial distance coordinate
$r$.

With $g_{rr}$ given by equation (\ref{eq:A}) the energy-density
turns out to be:

\begin{equation} \rho(r) = \frac{1}{8 \pi r^2} (1-\theta(r-r_h))
\end{equation}

Within the holostar's interior the mass-energy density follows an
inverse square law. Outside of the membrane, i.e. for $r > r_h$,
it is identical zero. Note, that $r_h$ must not necessarily be
finite.

In the following discussion the argument $(r
- r_h)$ of the $\theta$- and $\delta$-distributions will be omitted.

The radial and tangential pressures also follow from the metric:

\begin{equation} \label{eq:PrBH}
P_r = -\rho = -\frac{1}{8 \pi r^2} (1-\theta)
\end{equation}

\begin{equation}
P_\theta = P_\varphi = \frac{1}{16 \pi r_h} \delta
\end{equation}

$P_r$ is the radial pressure. It is equal in magnitude but
opposite in sign to the mass-density. $P_\theta$ denotes the
tangential pressure, which is zero everywhere, except for a
$\delta$-functional at the holostar's surface. The
"stress-energy-content" of the two principal tangential pressure
components in the membrane is equal to the gravitating mass $M$ of
the holostar.

In order to determine the principal pressures from the metric, the
time-coefficient of the metric $g_{tt}$ must be known. For the
holostar equation of state with $P_r = - \rho$ we have $g_{tt} =
1/g_{rr}$. Other equations of state lead to different
time-coefficients, and therefore different principal pressures.

Neither the particular form of the time-coefficient of the metric,
nor the particular form of the principal pressures are important
for the main results derived in this paper, which are based on
equilibrium thermodynamics, where time evolution is
irrelevant.\footnote{As long as the relevant time scale is long
enough, that thermal equilibrium can be attained.} The essential
assumptions are:
\begin{itemize}
\item spherical symmetry \item a radial
metric coefficient $g_{rr} = r/r_0$ \item a total energy density
$\rho = 1 / (8 \pi r^2)$ \item microscopic statistical
thermodynamics of an ideal gas of ultra-relativistic fermions and
bosons (in the context of the grand-canonical ensemble)
\end{itemize}

If the validity of Einstein's field equations with zero
cosmological constant is assumed, conditions two and three are
interchangeable.

In the following sections I assume that $r_0^2$ is nearly
constant, i.e. more or less independent of the size of the
holostar and comparable to the Planck area $A_{Pl} = \hbar$:

\begin{equation} \label{eq:r0}
{r_0}^2 = \beta {r_{Pl}}^2 = \beta \hbar
\end{equation}

This assumption will be justified later.

\section{\label{sec:fermion:simple}A simple derivation of the Hawking temperature and entropy}

The interior metric of the holostar solution is well behaved and
the interior matter-density is non-zero. The solution is static:
The matter appears to exert a radial pressure preventing further
collapse to a point singularity. However, the solution gives no
direct indication with respect to the state of the interior matter
and the origin of the pressure.

In this section I will discuss a very simple model for the
interior matter state of the holostar, which is able to explain
many phenomena attributed to black holes. Let us assume that the
interior matter distribution is dominated by ultra-relativistic
weakly interacting fermions and the pressure is produced by the
exclusion principle. Due to spherical symmetry the mean momentum
of the fermions $p(r)$ and their number density per proper volume
$f n(r)$  will only depend on the radial distance coordinate $r$.
$f$ denotes the effective number of degrees of freedom of the
fermions. For ultra-relativistic fermions the local energy-density
will be given by the product of the number density of the fermions
and their mean momentum. This energy density must be equal to the
interior mass-energy density of the holostar:

\begin{equation} \label{eq:rho_fnp}
\rho = p(r) f n(r) = \frac{1}{8 \pi r^2}
\end{equation}

If the fermions interact only weakly, their mean momenta can be
estimated by the exclusion principle:

\begin{equation} \label{eq:p3n}
p(r)^3 \frac{1}{n(r)} = (2 \pi \hbar)^3
\end{equation}

These two equations can be solved for $p(r)$ and $n(r)$:

\begin{equation} \label{eq:plocr}
p(r) = \frac{\hbar^{\frac{3}{4}}
\pi^{\frac{1}{2}}}{f^{\frac{1}{4}}} \frac{1}{r^{\frac{1}{2}}}
\end{equation}

\begin{equation} \label{eq:nlocr}
f n(r) = \frac{f^{\frac{1}{4}}}{\hbar^{\frac{3}{4}} 8
\pi^{\frac{3}{2}}} \frac{1}{r^{\frac{3}{2}}}
\end{equation}

The mean momenta of the fermions within the holostar fall off from
the center as $1/r^{1/2}$ and the number density per proper volume
with $1/r^{3/2}$. Similar dependencies, however without definite
factors, have already been found in \cite{petri/hol} by analyzing
the geodesic motion of the interior massless particles in the
holostar-metric. It is remarkable, that equilibrium thermodynamics
combined with the uncertainty principle gives the same results as
the geodesic equations of motion. This is not altogether
unexpected. In \cite{Jacobson} it has been shown, that the field
equations of general relativity follow from thermodynamics and the
Bekenstein entropy bound \cite{Bekenstein/81}.

The momentum of the fermions at a Planck-distance $r = r_{Pl} =
\sqrt{\hbar}$ from the center of the holostar is of the order of
the Planck-energy $E_{Pl}=\sqrt{\hbar}$. It is also interesting to
note, that for both quantities $p(r)$ and $n(r)$ the number of
degrees of freedom $f$ can be absorbed in the radial coordinate
value $r \rightarrow \sqrt{f} \, r$, so that $p$ and $n$
effectively depend on $\sqrt{f} \, r$. We will see later that the
square root of $f$ plays an important role in the scaling of the
fundamental length parameter $r_0$.\footnote{Perhaps it would be
better to reformulate the above statement, by saying that the
fundamental area $4 \pi r_0^2$ scales with $f$.}

From (\ref{eq:plocr}) one can derive the following momentum-area
law for holostars, which resembles the Stefan-Boltzmann law for
radiation from a black body:

\begin{equation} \label{eq:p4r2}
p(r)^4 r^2 f = \hbar^3 \pi^2
\end{equation}

Note that this law not only refers to the holostar's surface ($r
= r_h$) but is valid for any concentric spherical surface of
radius $r$ within the holostar. Therefore it is reasonable to
assume that the holostar has a well defined interior temperature
$T(r)$ proportional to the mean momentum $p(r)$:

\begin{equation} \label{eq:pT}
p(r) = \sigma T(r)
\end{equation}

$\sigma$ is a constant factor. We will see later, that it is related to
the entropy per particle.

The local surface temperature of the holostar is given by:

\begin{equation}
T(r_h) = \frac{p(r_h)}{\sigma} = \frac{\hbar^{\frac{3}{4}}
\pi^{\frac{1}{2}}}{\sigma f^{\frac{1}{4}}} \frac{1}{\sqrt{r_h}}
\end{equation}

The surface redshift $z$ is given by:

\begin{equation} \label{eq:z}
z = \frac{1}{\sqrt{g_{tt}(r_h)}} = \sqrt{g_{rr}(r_h)}=
\sqrt{\frac{r_h}{r_0}}
\end{equation}

where $g_{tt}(\infty) = 1$ is assumed.

The local surface temperature can be compared to the Hawking
temperature of a black hole. The Hawking temperature is measured
at infinity. Therefore the red-shift of the radiation emitted from
the holostar's surface with respect to an observer at spatial
infinity has to be taken into account, by dividing the local
temperature at the surface by the gravitational red shift factor
$z$. With $g_{rr}(r_h) = {r_h}^{1/2} ({\beta \hbar})^{-1/4}$  we
find:

\begin{equation} \label{eq:T_inf}
T_{\infty}= \frac{T(r_h)} {\sqrt{g_{rr}(r_h)}}=
\frac{\pi^{\frac{1}{2}}}{\sigma}
\left(\frac{\beta}{f}\right)^{\frac{1}{4}} \frac{\hbar}{r_h}
\end{equation}

The surface-temperature measured at infinity has the same
dependence on the gravitational radius $r_h$ as the Hawking
temperature, which is given by:

\begin{equation} \label{eq:T_Hawking}
T_{H} = \frac{\hbar}{4 \pi r_h} = \frac{\hbar}{8 \pi M}
\end{equation}

We get the remarkable result, that - up to a possibly different
constant factor - the Hawking temperature of a spherically
symmetric black hole and the respective temperature of the
holostar at infinity are equal.

As the Hawking temperature of a black hole only depends on the
properties of the exterior space-time, and the exterior
space-times of a black hole and the holostar are equal (up to a
small Planck-sized region outside the horizon), it is reasonable
to assume that the Hawking temperature should be the true
temperature of a holostar measured at spatial infinity. With this
assumption, the constant $\sigma$ can be determined by setting the
temperatures of equations (\ref{eq:T_inf}) and
(\ref{eq:T_Hawking}) equal:

\begin{equation} \label{eq:s} \label{eq:beta}
\sigma = \left(\frac{\beta}{f}\right)^{\frac{1}{4}} 4 \pi^{\frac{3}{2}}
\end{equation}

The total number of fermions within the holostar is given by the
proper integral over the number-density:

\begin{equation} \label{eq:Nint}
N = \int{f n(r) dV}
\end{equation}

$dV$ is the proper volume element, which can be read off from the
metric:

\begin{equation} \label{eq:dV}
dV = 4 \pi r^2 \sqrt{g_{rr}} dr = 4 \pi r^{\frac{5}{2}}
\left({\beta \hbar} \right)^{-\frac{1}{4}} dr
\end{equation}

Integration over the total interior volume of the holostar gives:

\begin{equation} \label{eq:Nclassic}
N = \left(\frac{f}{\beta}\right)^{\frac{1}{4}} \frac{1}{4
\pi^{\frac{3}{2}}} \frac{\pi r_h^2}{\hbar} = \frac{1}{\sigma}
\frac{A}{4 \hbar} = \frac{S_{BH}}{\sigma}
\end{equation}

$S_{BH}$ is the Bekenstein-Hawking entropy for a spherically
symmetric black hole with horizon surface area $A$.

Therefore the number of fermions within the holostar is
proportional to its surface area and thus proportional to the
Hawking entropy. This result is very much in agreement with the
holographic principle, giving it quite a new and radical
interpretation: The degrees of freedom of a highly relativistic
self-gravitating object don't only "live on the surface", the
object contains a definite number of particles\footnote{This is
not true for the Einstein-Maxwell vacuum black hole solutions with
event horizon. Due to the nature of the event horizon and its
accompanying singularity the number and nature of the particles
within a black hole - or rather gone into the black hole - is
indefinite.} and their total number is proportional to the
object's surface area, measured in units of the Planck-area,
$A_{Pl} = \hbar$. This result is an immediate consequence of the
interior metric $g_{rr} \propto r$, the energy-momentum relation
for relativistic particles $E = p$ and the exclusion principle. It
can be easily shown, that for any other spherically symmetric
metric, for example $g_{rr} \propto r^n$, the number of interior
(fermionic) particles is not proportional to the boundary
area.\footnote{Quite interestingly, the $N \propto r_h^2$ law can
be derived quite similar to the derivation of the
Chandrasekhar-limit for a white dwarf star, by assuming that the
sum of "gravitational energy" $E_{grav} \propto M / r$ and kinetic
energy $E_{kin} \propto N p$ has an extremum (in fact, a
maximum!). However, for the determination of the "gravitational
energy" of the star the proper radius $r_p = r \, (r/r_0)^{1/2}$
must be used instead of the radial coordinate $r$.}

From equation (\ref{eq:Nclassic}) we can see that $\sigma$ is the
entropy per particle. This allows a rough estimate of $\beta$: The
entropy of an ultra-relativistic particle should be of order unity
($\sigma \approx 3-4$). If we count the degrees of freedom of all
fermions in the Standard Model of particle physics (three
generations of quarks and leptons including the spin, color and
antiparticle degrees of freedom), their combined number is 90. The
number of bosonic degrees of freedom (8 gluons, 4 electro-weak
particles) is 24, disregarding the graviton and assuming the W-
and Z-bosons to be massless (above the energy of the electro-weak
phase transition). With the usual counting rule\footnote{This rule
is only true, when the chemical potential of fermions and bosons
is zero.}, weighting the fermionic degrees of freedom with $7/8$,
one gets $f =102.75$. Supersymmetry essentially doubles this
number. It is expected, that a unified theory will not vastly
exceed this number. For $\sigma = 3$ and $f = 256$ we find $4 \pi \beta
\simeq 1.06$. This justifies the assumption, that the fundamental
length parameter $r_0$ should be roughly equal to the
Planck-length.

By help of equation (\ref{eq:s}) the local temperature can be
expressed in terms of $\beta$ alone:

\begin{equation} \label{eq:Tlocal}
T(r) = \frac{\hbar^{\frac{3}{4}}}{4 \pi \beta^{\frac{1}{4}}}
\frac{1}{r^{\frac{1}{2}}} = \frac{1}{4 \pi} \frac{\hbar}{(r_0
r)^{\frac{1}{2}}}
\end{equation}

Note that $\beta$ depends explicitly on the (effective) number of
degrees of freedom $f$ of the ultra-relativistic particles within
the holostar via equation (\ref{eq:beta}). At the center of the
holostar all the fermion momenta are comparable to the Planck
energy, as can be seen from equation (\ref{eq:Tlocal}). All
fermions of the Standard Model of particle physics will be
ultra-relativistic. Quite likely there will be other fundamental
particles of a grand unified theory (GUT), as well as other
entities such as strings and branes. Thus, close to the holostar's
center the number of ultra-relativistic degrees of freedom will be
at its maximum and $\beta$ will be close to unity. The farther one
is distanced from the center, the lower the local temperature
gets. At $r \approx 10^6 km$ the electrons will become
non-relativistic.\footnote{Note that $r$ is not equal to the
proper distance to the center. The proper distance scales with $r
\sqrt{r/r_0}$. At $r \approx 10^9 m$ the proper distance to the
center is roughly $10^{30} m$, i.e. already vastly exceeding the
current (Hubble-) radius of the observable universe.} The only
particles of the Standard Model that remain relativistic at larger
radial positions will be the neutrinos. If all neutrinos are
massive, the mass of the lightest neutrino will define a
characteristic radius of the holostar, beyond which there are no
relativistic fermions contributing to the holostar's internal
pressure. If at least one of the neutrinos is massless, there will
be no limit to the spatial extension of a holostar.

Note, that the radial coordinate position at which the holostar's
interior radiation temperature is equal to the temperature of the
cosmic microwave background radiation, $T_{CMBR} = 2.725 \, K$,
corresponds to roughly $r \approx 10^{28} m\approx10^{12} \, ly$,
i.e. quite close to the radius of the observable universe. This is
just one of several coincidences, which point to the very real
possibility, that the holostar or a variant thereof actually might
serve as an alternative, beautifully simple model for the
universe. For a more detailed discussion including some definite
cosmological predictions, which are all experimentally verified
within an error of maximally 15 \% see \cite{petri/hol}.

Whenever the temperature within the holostar becomes comparable to
the mass of a particular fermion species, a phase transition is
expected to take place at the respective $r$-position. Such a
transition will lower the effective value of $f$, as one of the
particles "freezes" out.\footnote{The effective value of $f$ must
not necessarily change much. If we have a matter-antimatter
asymmetry and the chemical potential of the fermionic species that
"freezes out" is non zero (and higher than the temperature), there
is a chance that the effective value of $f$ remains nearly
constant: A significant part of the energy density of the frozen
out degrees of freedom will "survive" in the fermion with the high
chemical potential (a non-zero chemical potential "prefers" matter
over antimatter - or vice versa - and thus forbids the complete
annihilation of the fermionic species that is becoming
non-relativistic).} Whenever $f$ changes, either $\sigma$ or
$\beta$ must adjust due to equation (\ref{eq:beta}). The question
is, whether $\sigma$ or $\beta$ (or both) will change. Presumably
$\sigma$ will at least approximately retain a constant value: The
entropy per ultra-relativistic fermion, as well as the mean
particle momentum per temperature, appears to be a local property
which should not depend on the (effective) number of degrees of
freedom of the particles at a particular $r$-position.

Under the assumption that $\sigma$ is nearly constant, the ratio of
$\beta / f$ must be nearly constant as well, as can be seen
from equation (\ref{eq:s}). Whenever $f$ changes, $\beta$ will
adjust accordingly. Lowering the effective number of degrees of
freedom leads to a flattening of the temperature-curve, as heat
(and entropy) is transferred to the remaining ultra-relativistic
particles. At any radial position of a phase transition, where a
fermion becomes non-relativistic and annihilates with its
anti-particle, the temperature is expected to deviate from the
expression $T \propto 1/\sqrt{r}$. This is quite similar to what
is believed to have happened in the very early universe, when the
temperature fell below the electron-mass threshold and the
subsequent annihilation of electron/positron pairs heated up the
photon gas, keeping the temperature of the expanding universe
nearly constant until all positrons were destroyed.

If the "freeze-out" happens without significant heat and entropy
transfer to the remaining gas of ultra-relativistic particles,
such as when the particle that "freezes" out has an appreciable
non-zero chemical potential, the effective value of $f$ will
remain nearly constant, which would imply that $\beta$ be nearly
constant as well. In this case $\beta$ as well as $f$ would be
nearly constant universal quantities. There is evidence that this
might actually be the case.\footnote{See the discussion in section
\ref{sec:fermionBosonGas} and the related discussions in
\cite{petri/charge, petri/hol}.}

\section{\label{sec:fermionBosonGas}Thermodynamics of an ultra-relativistic fermion and boson gas}

In this section I will discuss a somewhat more sophisticated model
for the thermodynamic properties of the holostar.

As has been demonstrated in the previous section, if the holostar
contains at least one fermionic species, its properties very much
resemble the Schwarzschild vacuum black hole solution, when viewed
from the outside: Due to Birkhoff's theorem the external
gravitational field cannot be distinguished from that of a
Schwarzschild black hole. Its temperature measured at infinity is
proportional to the Hawking temperature.

Due to its non-zero surface-temperature and entropy the holostar
will gradually lose particles by emission from its surface. The
(exterior) time scale of this process will be comparable to the
Hawking evaporation time scale $\propto r_h^3$ (see for example
\cite{petri/hol}). The (exterior) time for a photon to travel
radially through the holostar is proportional to $r_h^2$.
Therefore even comparatively small holostars are expected to have
an evaporation time several orders of magnitude longer than their
interior relaxation time.

This allows us, with the possible exception of near Planck-size
holostars, to consider any spherical thin\footnote{With "thin" in
the present context we mean small compared to the radial
coordinate value $r$ at a particular position. Note however, that
even if $\delta r$ is small, the {\em proper} radial thickness of a
"thin" shell with radial extension $\delta r$ can be huge, because
the radial metric coefficient $g_{rr}$ scales with $r/r_0$.} shell
within the holostar's interior to be in thermal equilibrium with
its surroundings. Each shell can exchange particles, energy and
entropy with adjacent shells on a time scale much shorter than the
life-time of the holostar. Under these assumptions the
thermodynamic parameters within each shell can be calculated via
the grand canonical ensemble.

We mentally partition the holostar into a collection of
thin spherical shells. The temperature scales as $1/\sqrt{r}$ and
thus varies very slowly with $r$. For the chemical potential(s)
let us assume a slowly varying function with $r$ as
well.\footnote{Two natural choices present themselves: One is to to
assume a constant, possibly zero, chemical potential. The other is
to assume a chemical potential proportional to the temperature.}
This assumption will be justified later. Under these circumstances
the thickness of each shell $\delta r$ can be chosen such, that it
is large enough to be considered macroscopic, and at the same time
small enough, so that the temperature, pressure and chemical
potential(s) are effectively constant within the shell.

An accurate thermodynamic description has to take into account a
possible potential energy of position. For the holostar a
significant simplification arises from the fact, that the
effective potential $V_{eff}(r)$ for the radial motion of
massless, i.e. ultra-relativistic, particles is nearly constant,
as can be seen from the following discussion. The equations of
motion for ultra-relativistic particles within the holostar's
interior were given by \cite{petri/hol}:

\begin{equation} \label{eq:Eq:Veff}
\beta_r^2(r) + V_{eff}(r) = 1
\end{equation}

with

\begin{equation} \label{eq:EqMotion:r}
V_{eff}(r) = \frac{r_i^3}{r^3}
\end{equation}

and

\begin{equation} \label{eq:EqMotion:t}
\beta_\perp^2(r) = \frac{r_i^3}{r^3}
\end{equation}

$\beta_r(r)$ is the radial velocity of a photon, expressed as
fraction to the local velocity of light in the (purely) radial
direction. $\beta_\perp(r)$ is the tangential velocity of the
photon, expressed as a fraction to the local velocity of light in
the (purely) tangential direction. $r_i$ is the turning point of
the motion. For pure radial motion $r_i=0$.

We find that for pure radial motion the effective potential is
constant with $V_{eff}(r) = 0$. In the case of angular motion
($r_i \neq 0$) the effective potential approaches zero with
$1/r^3$, i.e. becomes nearly zero very rapidly, whenever $r$ is
greater than a few $r_i$. Therefore, to a very good approximation
we can regard the ultra-relativistic particles to move freely
within each shell. Their total energy will only depend on the
relativistic energy-momentum relation, not on the radial position.

With these preliminaries the grand canonical potential $\delta J$
of a small spherical shell of thickness $\delta r$ for a gas of
relativistic fermions at radial position $r$ will be given by:

$$
\delta J(r) = - T(r) \frac{f}{(2 \pi \hbar)^3} \delta V \int \int
\int d^3p \ln{(1+e^{-\frac{p- \mu(r)}{T(r)}})}
$$

$$ = - T^4 \delta V
\frac{f}{2 \pi^2 \hbar^3} \int_{z_{min}}^{z_{max}}{z^2
\ln{(1+e^{-z+\frac{\mu}{T}})} dz}$$

$z = p/T(r)$ is a dimensionless integration variable. Assuming
that we have a low and high energy cut-off, the integration over
$p$ ranges from $p_{min} \approx \hbar/(2 \pi r)$ to $p_{max}
\approx p_{Pl} = \sqrt{\hbar}$.\footnote{We will see later, that
these integration ranges can be replaced by 0 and $\infty$ to an
excellent approximation, so the question, whether there truly is a
low and/or high energy cutoff, as suggested by loop quantum
gravity, or whether there is no such cut-off, as advocated by
string-theory, is not relevant.} $\mu(r)$ is the chemical
potential at radial coordinate position $r$. $T(r)$ is the local
temperature at this position. $p_{PL}$ is the Planck-momentum,
which is equal to the Planck-energy in units $c=1$.

Note that even when the radial coordinate extension $\delta r$ of
the shell is small, the {\em proper} radial extension $\delta l =
(r/r_0)^{1/2} \delta r$ of the shell will become quite large
because of the large value of the radial metric coefficient in the
holostar's outer regions.

Knowing the results presented at the end of this section it is not
difficult to show that $z_{min} \approx 1 /N^{1/4}$ and $z_{max}
\approx N^{1/4}$, where $N$ is the number of particles in the
shell. With the exception of the central region of the holostar it
is possible to choose the radial extension of the shell such that
the number of particles within the shell, $N$, is macroscopic and
at the same time $T(r)$ and $\mu(r)$ are constant to a very good
approximation within the shell. For any holostar of macroscopic
dimensions the number of particles in its non-central shells will be
huge. Therefore the integration boundaries can be replaced to an
excellent approximation by zero and infinity.

\begin{equation} \label{eq:J0}
\delta J(r) = - T^4 \delta V \frac{f}{2 \pi^2 \hbar^3}
\int_0^\infty{z^2 \ln{(1+e^{-z+\frac{\mu}{T}})}dz}
\end{equation}

The proper volume of the shell $\delta V$ is given by the volume
element of equation (\ref{eq:dV}).

The ratio of chemical potential $\mu$ to local temperature $T$ is
assumed to be a very slowly varying function of $r$. In fact, we
will see later that this ratio is virtually independent of
$r$.\footnote{This might not be exactly true at the radial
coordinate position, where a phase transition takes place. I.e.
where a particle species undergoes a transition from relativistic
to non-relativistic motion.} The ratio $\mu/T$ will be denoted by
$u$, keeping in mind that $u$ might depend on $r$:

$$u = \frac{\mu(r)}{T(r)}$$

The integral in equation (\ref{eq:J0}) can be transformed to the
following integral by a partial integration:

\begin{equation} \label{eq:J}
\delta J(r) = - T^4 \delta V \frac{f}{2 \pi^2 \hbar^3} \frac{1}{3}
\int_0^\infty{z^3 n_F(z,u)dz}
\end{equation}

where $n_F$ is the mean occupancy number of the fermions:

\begin{equation} \label{eq:nF}
n_F(z, u) = \frac{1}{e^{z-u}+1} = \frac{1}{e^{\frac{p-\mu}{T}}+1}
\end{equation}

Knowing the grand canonical potential $\delta J$ the entropy
within the shell can be calculated:

\begin{equation} \label{eq:SF}
\delta S(r) = -\frac{\partial (\delta J)}{\partial T} = \frac{f}{2
\pi^2 \hbar^3} T^3 \delta V \left(\frac{4}{3} Z_{F,3}(u) - u
Z_{F,2}(u)\right)
\end{equation}

By $Z_{F,n}$ the following integrals are denoted:

\begin{equation} \label{eq:ZF}
Z_{F,n}(u) = \int_0^\infty{z^n n_F(z,u)dz}
\end{equation}

Such integrals, which commonly occur in the evaluation of
Feynman-integrals in QFT, can be evaluated by the poly-logarithmic
function $Li_n(z)$:

\begin{equation} \label{eq:ZF:Li}
Z_{F,n}(u) = -\Gamma(n+1) \, Li_{n+1}(-e^u)
\end{equation}

with $n+1>0$ and

\begin{equation} \label{eq:polylog:f}
Li_{n}(z) = \sum_{k=1}^\infty{\frac{z^{k}}{k^n}}
\end{equation}

For the derivation of the entropy the following identity has been
used, which is easy to derive from the power-expansion of
$Li_n(z)$.

\begin{equation}
\frac{\partial {Z_{F,3}(u)}}{\partial x} = 3 Z_{F,2}(u)
\frac{\partial u}{\partial x}
\end{equation}

The pressure in the shell is given by:

\begin{equation} \label{eq:PF}
P(r) = -\frac{\partial (\delta J)}{\partial (\delta V)} =
\frac{f}{2 \pi^2 \hbar^3} T^4 \frac{Z_{F,3}(u)}{3}
\end{equation}

The total energy in the shell can be calculated from the grand
canonical potential via:

\begin{equation} \label{eq:EF}
\delta E(r) = \delta J -\left(T \frac{\partial}{\partial T} + \mu
\frac{\partial}{\partial \mu}\right)\delta J = \frac{f}{2 \pi^2
\hbar^3} T^4 \delta V Z_{F,3}(u)
\end{equation}

The total number of particles within the shell is given by:

\begin{equation} \label{eq:NF}
\delta N(r) = -\frac{\partial (\delta J)}{\partial \mu}=\frac{f}{2
\pi^2 \hbar^3} T^3 \delta V Z_{F,2}(u)
\end{equation}

The total energy per fermion within the shell is proportional to
$T$, as can be seen by combining equations (\ref{eq:EF},
\ref{eq:NF}):

\begin{equation} \label{eq:EF_NF}
\epsilon = \frac{\delta E}{\delta N} =
\frac{Z_{F,3}(u)}{Z_{F,2}(u)} \, \, T(r)
\end{equation}

$\epsilon$ only depends indirectly on $r$ via $u$. We will see
later that $u$ is essentially independent of $r$, so that the mean
energy per particle is proportional to the temperature with nearly
the same constant of proportionality at any radial position $r$.

The entropy per particle within the shell can be read off from
equations (\ref{eq:SF}, \ref{eq:NF}):

\begin{equation} \label{eq:SF_NF}
\sigma = \frac{\delta S}{\delta N} = \frac{4}{3}
\frac{Z_{F,3}(u)}{Z_{F,2}(u)} - u
\end{equation}

Again, $\sigma$ only depends on $r$ via $u$.

The calculations so far have been carried through for fermions. It
is likely, that the holostar will also contain bosons in thermal
equilibrium with the fermions. The equations for an
ultra-relativistic boson gas are quite similar to the above
equations for a fermion gas. We have to replace:

\begin{equation} \label{eq:Replace:nB}
n_F(z,u) \rightarrow n_B(z,u) = \frac{1}{e^{z-u}-1}
\end{equation}

\begin{equation} \label{eq:Replace:zB}
Z_{F,n} \rightarrow Z_{B,n} = \int_0^\infty{z^n n_B(z,u) dz}
\end{equation}

with $n+1>0$ and

\begin{equation} \label{eq:polylog:b}
Z_{B,n} = \Gamma(n+1) \, Li_{n+1}(e^u)
\end{equation}

Let us assume that the fermion and boson gases are only weakly
interacting. In such a case the extrinsic quantities, such as
energy and entropy, can be simply summed up. The same applies for
the partial pressures.

The number of degrees of freedom of fermions and bosons can
differ. The fermionic degrees of freedom will be denoted by $f_F$,
the bosonic degrees of freedom by $f_B$. In general, the different
particle species will have different values for the chemical
potentials. There are some restraints. Bosons cannot have a
positive chemical potential, as $Z_{B,n}(u)$ is a complex number
for positive $u$. Photons and gravitons, in fact all massless
gauge-bosons, have a chemical potential of zero, as they can be
created and destroyed without being restrained by a
particle-number conservation law.

We are however talking of a gas of ultra-relativistic particles.
In this case particle-antiparticle pair production will take place
abundantly, so that we also have to consider the antiparticles.
The chemical potentials of particle and anti-particle add up to
zero: $\mu + \overline{\mu} = 0$. As bosons cannot have a positive
chemical potential, the chemical potential of any
ultra-relativistic bosonic species must be zero, i.e. $\mu_B =
\overline{\mu_B} = 0$, whenever the energy is high enough to
create boson/anti-boson pairs. This restriction does not apply to
the fermions, which can have a non-zero chemical potential at
ultra-relativistic energies, as both signs of the chemical
potential are allowed. So for ultra-relativistic fermions we can
fulfill the relation $\mu_F + \overline{\mu_F} = 0$ with non-zero
$\mu_F$.

For the following calculations it is convenient to use the ratio
of the chemical potential to the temperature $u = \mu / T$ as the
relevant parameter, instead of the chemical potential itself . If
the number of degrees of freedom of fermions and bosons
respectively, i.e. $f_F$ and $f_B$ is known, there are only two
undetermined parameters in the model, $u_F$ and
$\beta$.\footnote{More generally, if the chemical potentials of
the different particle species are different, one needs n-1
additional functional relations between  the n independent
chemical potentials.}

In order to determine $u_F$ and $\beta$ one needs two independent
relations. These can be obtained by comparing the holostar
temperature and entropy to the Hawking temperature and entropies
respectively.

Alternatively $u_F$ can be determined without reference to the
Hawking temperature law, solely by a thermodynamic argument. It is
also possible to determine $\beta$ by a theoretical argument as
proposed in \cite{petri/charge}.

The thermodynamic energy of a shell consisting of an
ultra-relativistic ideal fermion and boson gas is given by:

\begin{equation} \label{eq:E_th}
\delta E_{th} = \frac{F_E} {2 \pi^2 \hbar^3}  \delta V T^4
\end{equation}

with

\begin{equation}
F_E(u_F) = f_F ( Z_{F,3}(u_F) + Z_{F,3}(-u_F)) + 2 f_B Z_{B,3}(0)
\end{equation}

with the identities of the polylog-function and with $Z_{B,3}(0) =
\pi^4 / 15$ one can express $F_E$ as a quadratic function of
$u_F^2/\pi^2$ \cite{petri/asym}:

\begin{equation} \label{eq:FE:u2}
F_E(u_F) = 2 f_F \frac{\pi^4}{15} \left(\frac{15}{8}\left(1 +
\frac{\pi^2}{u_F^2} \right)^2 + \frac{f_B}{f_F} -1 \right)
\end{equation}

We take the convention here, that $f_F$ and $f_B$ denote the
degrees of freedom of one particle species, including particle and
antiparticle. With this convention a photon gas ($g=2$) is
described by $f_B = 1$ (There are two photon degrees of freedom
and the photon is its own anti-particle). All other particle
characteristics, such as helicities, are counted extra. The total
number of the degrees of freedom in the gas, i.e. counting
particles and anti-particles separately, will be given by

\begin{equation}
f = 2(f_F + f_B)
\end{equation}

The total energy of the holostar solution is given by the
proper integral over the mass density. The proper energy of the
shell therefore is:

\begin{equation} \label{eq:E_BH}
\delta E_{BH} = \rho \delta V = \frac{\delta V}{8 \pi r^2} =
\frac{1}{2} (\beta \hbar)^{-\frac{1}{4}} r^{\frac{1}{2}} \delta r
\end{equation}

Setting the two energies equal gives the local temperature within
the holostar:

\begin{equation} \label{eq:TlocTherm}
T^4 = \frac{\pi \hbar^3}{4 F_E r^2 }
\end{equation}

Thus we recover the $1/\sqrt{r}$-dependence of the local
temperature, at least if $F_E$ is constant.

$F_E$ is a function of $f_F$, $f_B$ and $u_F$. We will see later,
that $u_F$ only depends on the ratio of $f_F$ and $f_B$. Therefore
in any range of $r$-values where the number of degrees of freedom
of the ultra-relativistic particles (or rather their ratio)
doesn't change, the local temperature as determined by equation
(\ref{eq:TlocTherm}) will not deviate from an inverse square root
law.

If the temperature of equation (\ref{eq:TlocTherm}) is inserted
into equation (\ref{eq:PF}), the thermodynamic pressure is derived
as follows:

$$P(r) = \frac{1}{24 \pi r^2} = \frac{\rho}{3}$$

This is the equation of state for an ultra-relativistic gas, as
expected. Note, that the pressure doesn't exactly agree with the
pressure of the holostar solution, although it is encouraging that
the thermodynamic pressure at least has the right magnitude and
$r$-dependence. In fact, the magnitude of the thermodynamic
pressure is quite what is expected, when one takes into account
that the thermodynamic derivation above is ignorant of the
pressure anisotropy, treating all volume changes on an equal
footing, independent of the direction of the change: Within the
holostar the two tangential pressure components are zero, whereas
the radial pressure is equal to $-1/(8 \pi r^2)$. Therefore the
"averaged" pressure over all three spatial dimensions is $-1/(24
\pi r^2)$.

However, the thermodynamic pressure and the holostar pressure have
opposite signs. This discrepancy cannot be resolved in the simple
model discussed in this paper. This is not totally unexpected: The
holostar is a {\em string} solution. Yet the thermodynamic model
discussed in this paper assumes that the interior matter consists
{\em exclusively} out of {\em radiation}. While it is likely that
the interior matter will contain a significant radiation
contribution at high temperatures, it is unrealistic to assume
that the dominant matter-type in the holographic solution -
strings - is completely absent.\footnote{The inverse reasoning,
that {\em all} matter in the holostar solution must be strings due
to the interior string equation of state is incorrect. It is easy
to construct the stress-energy tensor of a string gas by
superposition of the stress-energy tensor for "normal" radiation
and an equal vacuum contribution. Furthermore one can construct
the interior stress-energy tensor of the holostar solution by
superimposing the stress-energy tensor for a spherically symmetric
charge-distribution with a vacuum contribution. See
\cite{petri/asym} or \cite{petri/charge} for more details.}

Note also, that for the derivation of the main results of this
paper, the particular form of the pressure is not essential, as
was pointed out in section \ref{sec:holo:intro}. Hopefully a
satisfactory explanation for the discrepancy can be found in the
future. A full explanation most likely will require a better
understanding of the "string-nature" of matter not only at high,
but also at low energies.

Another approach to resolve the problem is to replace the
holostar-solution with a somewhat more general, yet similar
solution to the field equations. For example, the equation of
state could be modified (slightly), such as $P_r = (-1 +
\delta(r)) \rho$. Even with $\delta(r) = const$ one gets a
significant tangential pressure component, as can be seen in
\cite{petri/bh}. Other modifications are thinkable. Yet one should
be reluctant to modify the condition $g_{rr} = r/r_0$, as the
results presented in this paper rely crucially on this condition.

\subsection{Comparing the holostar's thermodynamic temperature
and entropy to the Hawking result}

By inserting the temperature derived in equation
(\ref{eq:TlocTherm}) into equation (\ref{eq:SF}) we get the
following expression for the thermodynamic entropy within the
shell:

\begin{equation} \label{eq:Sshell}
\delta S(r) = \left(\frac{F_E}{4 \pi \beta}\right)^{\frac{1}{4}}
 \frac{F_S}{F_E} \frac{r \delta r}{\hbar}
\end{equation}

with

\begin{equation} \label{eq:FS}
F_S(u_F) = f_F \left( \frac{4}{3}  \{Z_{F,3}(u_F)\} - u_F
[Z_{F,2}(u_F)] \right) + 2 f_B \frac{4}{3}  \left( Z_{B,3}(0)
\right)
\end{equation}

We have used commutator [] and anti-commutator \{\} notation in
order to render the above relation somewhat more compact.

Using the identities for the polylog function it is possible to
express the above relation as a quadratic function of the variable
$u_F^2 / \pi^2$.

\begin{equation}
F_S = \frac{4}{3} F_E(u_F) - f_F \frac{\pi^4}{3} \frac{u_F^2}{\pi^2} \left(1 + \frac{u_F^2}{\pi^2} \right)
\end{equation}

with $F_E$ is given by equation (\ref{eq:FE:u2})

By comparing the temperature (\ref{eq:TlocTherm}) and the entropy
(\ref{eq:Sshell}) of the holostar solution derived in the context
of our simple model to the Hawking entropy and temperature, two
important relations involving the two unknown parameters of the
model $u_F$ and $\beta$ can be obtained.

We have already seen in section \ref{sec:fermion:simple} that the
holostar's temperature at infinity is proportional to the Hawking
temperature. As can be seen from equation (\ref{eq:TlocTherm})
this general result remains unchanged in the more sophisticated
thermodynamic analysis, as long as the quantity $F_E(u_F, f_F,
f_B)$ can be considered to be nearly constant. We will see later,
that the value of $u_F$ only depends on the ratio $f_B/f_F$, so
that $F_E = const$ whenever the number of fermionic and bosonic
degrees of freedom don't change. In order to determine $F_E$ we
can set the temperature at the holostar's surface equal to the
blue shifted Hawking temperature at the holostar's surface, which
can be obtained by multiplying the Hawking temperature (at
infinity) with the red-shift factor $z$ of the surface given in
Eq. (\ref{eq:z}). We find:

\begin{equation} \label{eq:TBH4}
T^4 = {T_{BH}}^4 \,  z^4 = \frac{\hbar^4}{2^8 \pi^4 {r_h}^4} \cdot
\frac{r_h^2}{\beta \hbar} = \frac{\hbar^3}{2^8 \pi^4 \beta
{r_h}^2}
\end{equation}

Comparing this to equation (\ref{eq:TlocTherm}) we find:

\begin{equation} \label{eq:FE:beta}
\frac{F_E}{4 \pi \beta} = (2 \pi)^4
\end{equation}

This is an important result. It relates the fundamental area $4
\pi r_0^2 = 4 \pi \beta \hbar$ to the thermodynamic parameters of
the system, i.e. the number of degrees of freedom and the chemical
potential of the fermions.

Another important relation  is the ratio $F_S/F_E$ in the interior
holostar space-time, which can be obtained by comparing the
Hawking entropy of a black hole with thermodynamic entropy of the
holostar's interior constituent matter.

The entropy of the holostar can be calculated by integrating
equation (\ref{eq:Sshell}). We will assume that $F_E / \beta =
const$, as follows from equation (\ref{eq:FE:beta}), and that
$F_S/F_E = const$, which will be justified shortly. If this is the
case, the integral can be performed easily:

\begin{equation}
S = \int_0^{r_h}{\delta S(r) dV} = \left(\frac{F_E}{4 \pi
\beta}\right)^{\frac{1}{4}} \frac{1}{2 \pi}
 \frac{F_S}{F_E} \frac{A}{4 \hbar}
\end{equation}

with

$$A = 4 \pi {r_h}^2$$

Setting this equal to the Hawking entropy, $S_{BH} = A/(4 \hbar)$,
and using equation (\ref{eq:FE:beta}) we find the important
result:

\begin{equation} \label{eq:FS:FE}
\frac{F_S}{F_E} = 1
\end{equation}

Writing out the above equation we get:

\begin{equation} \label{eq:ImplicituF:0}
\frac{f_F\left(\frac{4}{3} \{Z_{F,3}(u_F)\} - u_F
[Z_{F,2}(u_F)]\right)+ 2 f_B\left(\frac{4}{3}
Z_{B,3}(0)\right)}{f_F \{Z_{F,3}(u_F)\} + 2 f_B Z_{B,3}(0)} = 1
\end{equation}

which can be simplified to

\begin{equation} \label{eq:ImplicituF}
\frac{u_F [Z_{F,2}(u_F)]}{\{Z_{F,3}(u_F) \} + 2 \frac{f_B}{f_F}
Z_{B,3}(0)} = \frac{1}{3}
\end{equation}

Using the identities for the polylog function one can reduce the
above equation to a very simple quadratic equation in the variable
$u_F^2 / \pi^2$:

\begin{equation} \label{eq:uF:expl}
\left(1 + \frac{u_F^2}{\pi^2} \right) \left(1 - 3
\frac{u_F^2}{\pi^2} \right) + \frac{8}{15} \left(\frac{f_B}{f_F}
-1\right) = 0
\end{equation}

The important message, which can be seen
already from equation (\ref{eq:ImplicituF}) is, that whenever the
bosonic and fermionic degrees of freedom - or rather their ratio
$f_B/f_F$ - is known, $u_F$ can be calculated. Knowing $u_F$,
$\beta$ can be determined via (\ref{eq:FE:beta}). Thus the two
relations (\ref{eq:FE:beta}, \ref{eq:FS:FE}) allow us to determine
all free parameters of the model, whenever the number of particle
degrees of freedom, $f_F$ and $f_B$ are known.

\subsection{An alternative derivation of the relation $F_S/F_E = 1$}

Before discussing the specifics of the thermodynamic model, I
would like to point out another derivation of equation
(\ref{eq:FS:FE}), which does not depend on the Hawking result.
This alternative derivation only depends on the following
fundamental thermodynamic relation

\begin{equation} \label{eq:SET}
\frac{\delta S}{\delta E} T = 1
\end{equation}

and on the fact, that the holostar's interior matter state is
completely rigid, i.e. the interior matter state at any particular
radial position depends only on $r$, but not on the overall size
of the holostar.

Consider a process, where an infinitesimally small spherical shell
of matter is added to the outer surface of the holostar. This
process doesn't affect the inner matter of the holostar, as the
interior matter-state of the holostar at a given radial
coordinated position $r$ does not depend in any way on the size of
the holostar or on any other global quantity. Therefore, when
adding a new layer of matter we don't have to consider any
interaction, such as heat-, energy- or entropy-transfer between
the newly added matter layer and the interior
matter.\footnote{This statement implicitly assumes, that we can
neglect the effect of the boundary membrane, which might be an
oversimplification. When we place a new layer of matter with
radial extension $dr$ at the former boundary $r$ of the holostar,
the boundary membrane moves throughout the newly added layer to
its new position at $r+dr$. It is not altogether clear, whether
this process is adiabatic. However, in section \ref{sec:membrane}
arguments are given, that the membrane has zero entropy, so that
the assumption, that the different initial and final states of the
membrane have no effect on the thermodynamics of the process,
seems not too far fetched.} It is an adiabatic process, for which
we can calculate the entropy-change of the whole system via
equation (\ref{eq:SET}). Let $r$ be the radial position of the
holostar's surface. The entropy of the newly added shell is given
by equation (\ref{eq:Sshell}), its energy by equation
(\ref{eq:E_BH}), and its temperature by equation
(\ref{eq:TlocTherm}). One finds that the thermodynamic relation
(\ref{eq:SET}) is only fulfilled, when $F_S = F_E$. We have
derived equation (\ref{eq:FS:FE}) only from thermodynamics.

\subsection{A closed formula for $u_F$ and some special cases}

The chemical potential per temperature $u_F$ can be determined by
finding the root of equation (\ref{eq:uF:expl}). The value of
$u_F$ depends only on the ratio of fermionic to bosonic degrees of
freedom.\footnote{and on the constant ratio $F_S/F_E$, which has
been shown to be unity for the interior holostar solution}. Let us
denote the ratio of the degrees of freedom by

\begin{equation}
r_f = \frac{f_B}{f_F}
\end{equation}

Then $u_F$ is given by:

\begin{equation} \label{eq:u2}
\frac{u_F^2}{\pi^2} = \frac{2}{3} \sqrt{1 + \frac{2}{5} (r_f-1)} -
\frac{1}{3}
\end{equation}

For $r_f=0$ (only fermions) we find the following result:

\begin{equation} \label{eq:uF:rf=0}
u_F = \pi \, \sqrt{\sqrt{\frac{4}{15}} - \frac{1}{3}}  = 1.34416
\end{equation}

For $r_f=1$ (equal number of fermions and bosons) we get:

\begin{equation} \label{eq:uF;rf=1}
u_F = \frac{\pi}{\sqrt{3}} = 1.8138
\end{equation}

From equation (\ref{eq:u2}) one can see that $u_F$ is a
monotonically increasing function of $r_f$. It attains its minimum
value, when there are no bosonic degrees of freedom, i.e. $f_B =
r_f = 0$. When the bosonic degrees of freedom vastly exceed the
fermionic degrees of freedom, $u_F$ can - in principle - attain
high values. For large $r_f$ we have $u_F \propto (r_f-1)^{1/4}$.
For all practical purposes one can assume that the number of
bosonic degrees of freedom is not very much higher than the number
of fermionic degrees of freedom. This places $u_F$ in the range
$1.34 < u_F < 3$.

It is important to notice, that equation (\ref{eq:ImplicituF})
only has a solution when the number of fermionic degrees of
freedom, $f_F$, is non-zero, whereas $f_B$ can take arbitrary
values for any non-zero $f_F$. Therefore at least one fermionic
(massless) particle species with a non-vanishing chemical
potential proportional to the local radiation temperature is
necessary, if the interior mass-energy-density $1 / (8 \pi r^2)$
of the holostar is to be in thermodynamic equilibrium.

\subsection{Thermodynamic relations, which are independent from the Hawking formula}

If $u_F$ is known, all thermodynamic quantities of the model, such
as $F_E(u_F)$ and $F_N(u_F)$ etc. can be evaluated. Note that in
order to determine $u_F$ we only needed the relation $F_E = F_S$,
whose derivation didn't require the Hawking temperature/entropy
relation. Yet in order to fix $\beta$ via equation
(\ref{eq:FE:beta}) we had to compare the holostar's temperature
(or entropy) to the Hawking-result. Therefore the particular
relation between $\beta$ and $F_E$ derived in equation
(\ref{eq:FE:beta}) is tied to the the validity of the Hawking
temperature formula.

Although there is no doubt that the Hawking temperature of a large
black hole must be inverse proportional to its mass\footnote{This
already follows from the Bekenstein-argument, that the entropy of
a black hole should be proportional to the surface of its event
horizon.}, the exact numerical factor has not yet been determined
experimentally and thus might be questioned. For example, the
Hawking entropy/temperature could be subject to a moderate
rescaling\footnote{There are two possible effects which could
influence the value $4 \hbar$ in the denominator of the Hawking
entropy-area formula $S = A / (4 \hbar)$. First $4 \hbar$ is a
"fundamental area". Its value depends on Newton's constant. It has
been speculated, that Newton's constant might undergo a (finite)
renormalization depending on the energy scale. Second, the
holostar's membrane isn't situated at the gravitational radius of
the holostar, but roughly a Planck coordinate distance outside.
Therefore the holostar's temperature at infinity might be slightly
lower (and its entropy higher) than the Hawking result, which
assumes an exterior vacuum space-time right up to the horizon. The
second effect should be quite negligible for large holostars.}, so
it is worthwhile to know what thermodynamic relations in the
interior holostar space-time are independent from the Hawking
formula. The following derivations only make use of equation
(\ref{eq:FS:FE}), i.e.  $F_E = F_S$.

Knowing $u_F$ from equation (\ref{eq:u2}) the entropy per
particle, $\sigma$, can be easily calculated by equations
(\ref{eq:SF}, \ref{eq:NF}):

\begin{equation} \label{eq:SperN}
\sigma = \frac{\delta S}{\delta N} = \frac{F_S}{F_N} =
\frac{F_E}{F_N}
\end{equation}

with

\begin{equation} \label{eq:FN}
F_N(u_F) = f_F (Z_{F,2}(u_F) + Z_{F,2}(-u_F))+ 2 f_B Z_{B,2}(0)
\end{equation}

The energy per relativistic particle is given by equations
(\ref{eq:E_th}, \ref{eq:N:FN}). We find, just as in the previous
section, that the mean particle energy per temperature is
constant\footnote{at least as long as $f_F$ and $f_B$ remain
constant} and equal to the mean entropy per particle:

\begin{equation} \label{eq:EperN}
\epsilon = \frac{\delta E}{\delta N} = \frac{F_E}{F_N} T = \sigma
T
\end{equation}

\begin{figure}[t]
\begin{center}
\includegraphics[width=12cm, bb=60 433 500 767]{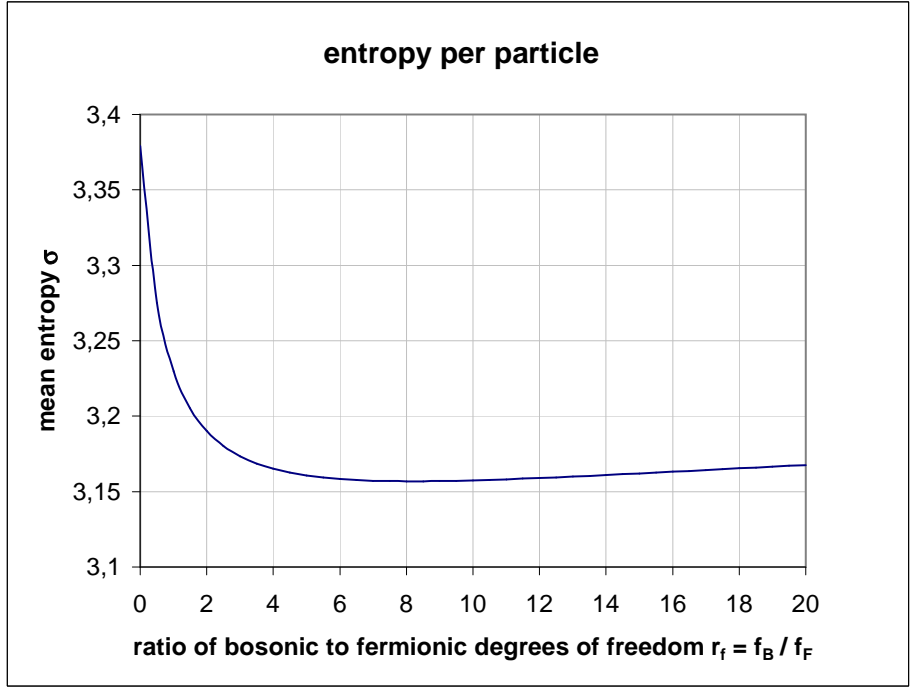}
\caption{\label{fig:epp}mean entropy per particle of the
ultra-relativistic fermions in the holographic solution as a
function of the ratio of bosonic to fermionic degrees of freedom
$r_f = f_B / f_F$}
\end{center}
\end{figure}

$\sigma$ only depends on the number of degrees of freedom of the
ultra-relativistic bosons and fermions in the model. In fact,
$\sigma$ only depends on the ratio $r_f = f_B/f_F$ and is a very
slowly varying function of this ratio. Figure \ref{fig:epp} shows
the dependence of $\sigma$ on $r_f$. In section \ref{sec:Tables}
the values of the entropy per particle $\sigma$, the ratio of
chemical potential to temperature of the fermions $u_F$, and other
interesting thermodynamic parameters are tabulated for several
values of $f_F$ and $f_B$.

The relation $\epsilon = \sigma T$, which relates the mean energy
per particle to the mean entropy times the local radiation
temperature can be viewed as the {\em fundamental thermodynamic
characteristic} of the holostar. Keep in mind that this relation
is only valid for the {\em mean} energy per particle and the {\em
mean} entropy per particle, evaluated with respect to all
particles. It isn't fulfilled for the bosonic and fermionic
species individually. In general, except for the special case $f_B
= 0$, we have $\epsilon_B \neq \sigma_B T$ and $\epsilon_F \neq
\sigma_F T$.

The relation $\epsilon = \sigma T$, which is equivalent to $F_S =
F_E$, has the remarkable side-effect, that the free energy is
identical zero in the holostar solution:

\begin{equation} \label{eq:F}
F = E - S T = N (\epsilon - \sigma T ) = 0
\end{equation}

Usually a closed system has the tendency to minimize it's free
energy, which is a compromise between minimizing it's energy and
maximizing it's entropy. The holostar is the prototype of a closed
system. It is a self-gravitating static solution to the Einstein
field equations. It's only form of energy-exchange with the outer
world is through Hawking-radiation, which is an utterly negligible
mode of energy-exchange for a large holostar. In this respect it
is remarkable that the holostar solution minimizes the free energy
to zero, e.g. the smallest possible value that a sensible measure
of energy in general relativity can have\footnote{In general
relativity the total energy is always positive}. One can speculate
on the basis of this result, whether the free energy in general
relativity might be more than a mere book-keeping device.

With the help of equation (\ref{eq:FS:FE}), but not using equation
(\ref{eq:FE:beta}), the entropy within the shell can be expressed
as:

\begin{equation} \label{eq:delta:S}
\delta S(r) = \left(\frac{F_E}{4 \pi
\beta}\right)^{\frac{1}{4}}\frac{r \delta r }{\hbar}
\end{equation}

If the total entropy of the holostar, i.e. the integral over the
entropy-contributions of the respective shells, is to be
proportional to the Hawking entropy of a black hole with the same
gravitational radius, $F_E / \beta$ must be constant. Integration
of equation (\ref{eq:delta:S}) gives the result:

\begin{equation} \label{eq:S:unfixed}
S = \frac{1}{2\pi} \left(\frac{F_E}{4 \pi \beta}\right)^{\frac{1}{4}} \frac{A}{4 \hbar}
\end{equation}

The Hawking result is reproduced, whenever:

\begin{equation} \label{eq:kappa}
\omega = \frac{1}{2 \pi} \left(\frac{F_E}{4 \pi
\beta}\right)^{\frac{1}{4}} = 1
\end{equation}

$\omega$, which depends on the ratio $F_E / \beta$, is the
constant of proportionality between the holostar entropy and the
Hawking entropy. Setting $\omega = 1$ is equivalent to equation
(\ref{eq:FE:beta}), which fixes $\beta$ with respect to the
Hawking temperature. If the Hawking entropy/temperature formula
have to be rescaled, $\omega$ is nothing else than the (nearly
constant) scale factor. Therefore let us express all thermodynamic
relations in terms of $\omega$.

The number of particles within the shell is given by equation
(\ref{eq:NF}), which is extended to encompass the bosonic degrees
of freedom:

\begin{equation} \label{eq:N:FN}
\delta N(r) = \frac{F_N}{F_E} \left(\frac{F_E}{4 \pi
\beta}\right)^{\frac{1}{4}} \frac{r \delta r}{\hbar} =
\frac{\omega}{\sigma} \frac{2 \pi r \delta r}{\hbar}
\end{equation}

The total number of particles is given by a simple integration,
assuming that $\omega = const$:

\begin{equation} \label{eq:N:holostar}
N = \left(\frac{F_E}{4 \pi \beta}\right)^{\frac{1}{4}} \frac{1}{2
\pi} \frac{1}{\sigma} \frac{A}{4 \hbar} = \frac{\omega}{\sigma}
\frac{A}{4 \hbar}
\end{equation}

Therefore, as derived in the previous section, the total number of
particles within the holostar is proportional to its surface area,
whenever $F_E / \beta = const$ and $\sigma = const$.

The temperature of the holostar at infinity is given by

\begin{equation} \label{eq:T:inf:unfixed}
T_{\infty} = T(r_h) \sqrt{g_{tt}(r_h)} = 2\pi \left(\frac{4 \pi
\beta}{F_E }\right)^{\frac{1}{4}} \frac{\hbar}{4 \pi r_h} =
\frac{1}{\omega} \frac{\hbar}{4 \pi r_h}
\end{equation}

Again, if we set $\omega = 1$ we get the Hawking temperature. The
important result is, that $\omega$ could in principle take on any
arbitrary (nearly constant) value. This is possible, because the
factor in the temperature is just the inverse as the factor in the
entropy. As is well known from black hole physics, any constant
rescaling of the Hawking entropy must necessarily rescale the
temperature such, that the product of temperature and entropy is
equal for the scaled and unscaled quantities, i.e. $S \, T$ must
be unaffected by the rescaling. This is necessary, because
otherwise the thermodynamic identity

$$\frac{\partial S}{\partial E} T = 1$$

would not be fulfilled in the exterior space-time. (In the
exterior space-time the energy $E$ is fixed and is taken to be the
gravitating mass $M = r_h/2$ of the black hole.)

As can be seen from equations (\ref{eq:S:unfixed},
\ref{eq:T:inf:unfixed}), entropy and temperature at infinity of
the holostar fulfill the rescaling condition. Furthermore, entropy
and temperature at infinity are exactly proportional to the
Hawking temperature and entropy. This result is not trivial. It
depends on the holostar metric, which has just the right value at
the position of the membrane, so that the temperature at infinity
scales correctly with respect to the entropy.

\subsection{Relating the local thermodynamic temperature to the Hawking temperature}

Now we are ready to set $\omega=1$, which gives us the desired
relation between $\beta$ and $F_E$, as already expressed in
equation (\ref{eq:FE:beta}).

With $\omega = 1$, the local thermodynamic temperature of any
interior shell can be expressed solely in terms of $\beta$. It
turns out to be equal to the expression in equation
(\ref{eq:Tlocal}) of the previous section:

\begin{equation} \label{eq:T4local}
T^4 = \frac{\hbar^3}{(4 \pi)^4 \beta} \frac{1}{r^2}
\end{equation}

or

$$T^4 A = \frac{1}{\beta} \left(\frac{\hbar}{(4 \pi)} \right)^3 = const$$

\section{A measurement of the Hawking temperature}

In the previous section the internal temperature of the holostar
has been derived by "fixing" it with respect to the Hawking
temperature. Although Hawking's calculations are robust and there
appears to be no reason, why the Hawking equation should be
modified - at least for large black holes\footnote{The only
ingredient in Hawking's derivation is the propagation of a quantum
field in the exterior vacuum space-time of a black hole. Both
concepts (quantum field in vacuum; exterior space-time of a black
hole) are very accurately understood.} - it has been speculated
whether the factor in the entropy-area law (or in the temperature
formula) might take a different value. A single measurement of the
Hawking temperature (or entropy) of a large black hole could
settle the question. However, with no black hole available in our
immediate vicinity and taking into account the extremely low
temperatures of even comparatively small black holes, there
appeared to be no feasible means to measure the Hawking entropy or
temperature of a black hole directly or indirectly.

It would be of high theoretical value, if the Hawking
temperature/entropy formula could be verified (or falsified) by an
explicit measurement. The holostar provides such a means.

For this purpose let us assume, that the Hawking temperature formula
were modified by a constant factor, i.e

\begin{equation}
T = \frac{1}{\omega} \frac{\hbar}{4 \pi r}
\end{equation}

where $\omega$ is a dimensionless factor, whose value can be
determined experimentally.

If we set the temperature of the holostar equal to the modified
Hawking temperature we get the following result for $F_E$:

\begin{equation}
\frac{F_E}{4 \pi \beta} = \left(2 \pi \omega \right)^4
\end{equation}

The local temperature within the holostar is then given by
equation (\ref{eq:TlocTherm}):

\begin{equation}
T^4 = \frac{\hbar^3}{2^8 \pi^4 \beta r^2 \omega^4} =
\frac{1}{\omega^4} \frac{\hbar^3}{2^5 \pi^3 \beta} \rho
\end{equation}

$\rho = 1/(8\pi r^2)$ is the total (local) energy density of the
matter within the holostar. The above equation can be solved for
$\omega$:

\begin{equation}
\omega^4 = \frac{\hbar^3} {2^5 \beta \pi^3} \frac {\rho} {T^4}
\end{equation}

The local radiation temperature $T$ and the total local energy
density $\rho$ within a holostar are both accessible to
measurement. Note that the local temperature within a holostar is
much easier to measure than its (Hawking) temperature at infinity:
The local interior temperature only scales with $1/\sqrt{M}$,
whereas the temperature at infinity scales with $1/M$. Therefore
even a very large holostar will have an appreciable interior local
radiation temperature, although its Hawking temperature at
infinity will be unmeasurable by all practical means.

In order to determine $\omega$ the value of $\beta$ need to be
known. In \cite{petri/charge} the following formula for $\beta$
has been suggested:

\begin{equation} \label{eq:beta:running}
\frac{\beta}{4} = \frac{\alpha}{2} + \sqrt{\left(
\frac{\alpha}{2}\right)^2 + \frac{3}{4}}
\end{equation}

$\alpha$ is the running value of the fine-structure constant,
which depends on the local energy scale. Note that the above
relation for $\beta$ has not been derived rigorously in \cite{petri/charge}, but was suggested by analogy, i.e. by extrapolating the (exact) formula derived for an extremely charged holostar to the rotating case. Angular momentum was introduced in rather straightforward way, giving the correct formula for a Kerr-Newman black hole in the macroscopic limit and the correct formula for the non-rotating case ($J=0$). The extrapolation formula then was applied to a {\em microscopic} object, a spin-1/2 extremely charged holostar of minimal mass, in order to obtain equation (\ref{eq:beta:running}). One must keep in mind though, that in principle there are several different choices, which give the correct macroscopic limit, but which might differ in their microscopic predictions.

We wan't to apply equation (\ref{eq:beta:running}) in order to derive the value of $r_0$ which determines the {\em interior} radial metric coefficient of a large holostar, $g_{rr} = r_0/r$. The implicit assumption which lies at the heart of equation (\ref{eq:beta:running}), is that $r_0^2 = \beta \hbar$ is a {\em universal} quantity, not dependent on the nature of the system in question and only - moderately - dependent on the energy-scale. It requires quite a leap of faith to do this. However, this assumption will be - at least partly - theoretically
justified by the discussion in section \ref{sec:asymmetry} and is
backed by some experimental evidence, as will be shown in the same
section.

If we assume $r_0^2 = \beta \hbar$ to be universal, it is not quite clear how the value of $\alpha$ should be interpreted when the local energy-densities and temperatures are high. In the original "derivation" of equation
(\ref{eq:beta:running}) $\alpha$ referred to the value of the fine-structure constant for a single charged spin-1/2 minimal mass holostar, which was identified with an electron {\em at rest}. The original "derivation" therefore refers to a low-energy situation, where the electromagnetic field is the only long-range field
besides gravity.

Due to the appearance of $\alpha$ in the formula for $r_0^2$ it is
suggestive to interpret $r_0$ as a running length scale, which
depends on the energy $E$ via $\alpha(E)$. This means that for
high temperatures $r_0$ is expected to increase with energy as a
function of $\alpha(E)$. This makes sense, because we have already
seen, that $r_0^2$ is proportional to the effective degrees of
freedom, which are also known to increase at high energies.
Therefore, if we treat $r_0^2$ as a universal quantity, the only
sensible way is to interpret $\alpha$ as the running value of the
relevant coupling constants depending on the energy scale.
Whenever the energy scale becomes comparable to the strong
interaction scale or the electro-weak unification scale, the other
coupling constants have to be properly accounted for.

With this interpretation, whenever $\alpha$ is small, such as for
the typical energies encountered today, it can be set to zero in
the above equation to a very good approximation, so that $\beta
\approx 4 \sqrt{3/4}$.

Let us now make the assumption, that we live in a large holostar.
In \cite{petri/hol} several observational facts have been
accumulated which suggest that such a claim is not too far
fetched. Then the local radiation temperature will be nothing else
than the microwave-background temperature and the total (local)
energy density will be the total matter density of the universe at
the present time (= present radial position). Both quantities have
been determined quite precisely in the recent past. With the
following value for the temperature of the microwave background
radiation

$$T_{CMBR} = 2.725 K$$

and with the total matter density determined from the recent
WMAP-measurements \cite{WMAP/cosmologicalParameters}

$$\rho = 0.26 \, \rho_{c} = 2.465 \cdot 10^{-27}
\frac{kg}{m^3}$$

and with $\beta$ determined from equation
({\ref{eq:beta:running}}) using the present (low energy) value of
the fine-structure constant, $\alpha$,

$$\beta = 3.479$$

we find:

\begin{equation}
\omega^4 = 1.0116
\end{equation}

or

$$\omega = 1.003$$

If we set the fine-structure constant to zero, i.e. $\beta = 4
\sqrt{3/4}$, the agreement is almost as good: $\omega = 1.004$.
The very high accuracy suggested in the above results is somewhat
deceptive. With $T$ known to roughly 0.1\% the error in $\omega$ will
be dominated by the uncertainty in $\rho$. A conservative estimate
for this uncertainty should be roughly 5\%. Taking the fourth
square root suppresses the relative error by roughly a factor of
four, so that the error in $\omega$ will be roughly 1\%.
Therefore, within the uncertainties of the determination of $\rho$
and $T$ the Hawking-entropy formula is reproduced to a remarkably
high degree of accuracy of roughly 1\%.

\section{\label{sec:Hawking:scaled}A scale invariant reformulation for the Hawking entropy}

In this section I briefly discuss some implications of a
(moderate) rescaling of the gravitational constant $G$ at high
energies, i.e. whether the formula for the entropy, particle
number, local temperature and temperature at infinity derived in
the last sections are invariant under a rescaling of $G$, or if
they can be reformulated in an invariant manner.

Before I do this, let me clarify my own position on this issue. If
one believes in the field equations of general relativity, which
are independent from a particular energy scale, then one cannot
possibly come to think that $G$ should be an energy-dependent
quantity. Much of the beautiful geometric interpretation of the
field equations were lost, if $G$ were variable. In effect, a
variable $G$ would completely change the internal structure of the
theory. The constancy of $G$ is an essential requirement in the
Einstein equations. Yet these equations have only been verified at
low energies. There is a common understanding, that at high
energies the field equations might have to be modified. The best
suited candidate theory for such a modification is string-theory.
Does string-theory require $G$ to be running? I doesn't
necessarily appear so. $G$, or rather a combination of $G$ and the
other constants ($c$, $\hbar$) with the dimension of length is
needed to define the dimensionless string coupling. Yet many
string-theorists tend to view the string-coupling as a quantity
that does not run. As long as there is no hard theoretical
evidence for a running $G$, I prefer the purist position, that $G$
is a true constant of nature, as the other two constants $c$ and
$\hbar$.

Yet there is no proof for this position, so it is instructive to
reflect on the possible implications that a running $G$ might
have.

The notation that was used so far is somewhat inconvenient for the
purpose of this section, because $G = 1$ was assumed in the
previous sections. It is not difficult to re-introduce $G$. Any
occurrence of $\hbar$ has to be replaced by $\hbar G$ and enough
powers of $\hbar$ have to be inserted into the equations, so that
the dimensions come out correct. The convention $c=1$ will be
retained, i.e. length and time are interchangeable. $G$ then has
dimensions $[m/kg]$, i.e. allows us to express any mass as a
length and vice versa. $\hbar$ is complementary to $G$. Its
dimensions are $[m \, kg]$, so that via $\hbar$ any length can be
expressed as an inverse mass (or energy) and vice
versa.\footnote{There appears to be some sort of duality between
$\hbar$ and $G$: In a certain sense $\hbar$ and $G$ are
complimentary with respect to transforming a length- into an
energy scale. Whereas $\hbar$ transforms length into inverse
energy, $G$ transforms length into energy directly.}

In contrast to $G$, which might behave like a running coupling
constant, $c$ should be thought of as a true constants of nature,
independent of the energies of the respective
interactions.\footnote{The constancy of $c$ is linked to one of
the most important space-time symmetries, i.e. Lorentz-invariance,
which is one of the basic building blocks of classical and quantum
physics. Nobody should give up Lorentz-invariance lightly.}
$\hbar$ also should be viewed as a true constant of nature,
independent of the energy scale.\footnote{$\hbar$ arises from
rotational symmetry which dictates the quantization of angular
momentum in half-integer steps of $\hbar$. If $\hbar$ were
variable, we would have a severe problem with the conservation law
for angular momentum.}

From $G$ and $\hbar$ two important combinations can be formed. In
units $c=1$ the quantity $\hbar G$ has the dimensions of area,
i.e. can be considered as a fundamental area. Furthermore the
quantity $\hbar^2 / (\hbar G) = \hbar / G$ has dimensions of
energy (or mass) squared, so that $\hbar G$ can alternatively be
regarded as an inverse energy squared.\footnote{Note that the
cross-sectional area $\sigma$ of the high energy interactions
between particles in the standard model follows an inverse square
law in the energy, i.e. $\sigma \propto \hbar^2/E^2$ when the
rest-masses of the particles are negligible. The cross-sectional
area of a black hole (or holostar), however, scales with $E^2$, so
there is some sort of duality as well. This has some resemblance
to string theory, where the winding modes and the vibrational
modes also have a different (proportional vs. inverse
proportional) dependencies on the radius of the circular
dimension. In string theory, a state with a particular winding and
vibration number is indistinguishable from a state, where winding
and vibration number are interchanged.}

The equations derived in the previous section depend on the value
of the gravitational constant $G$, which enters into the
definition of the fundamental area $\hbar \rightarrow \hbar G
/c^3$. The formula are not necessarily invariant under a rescaling
of $G$. For example, the entropy is given by $S = A / (4 \hbar
G)$. Whenever the value of $G$ is modified, the entropy of the
system changes due to the change in the fundamental area (unless
the measurement process of the area $A$ somehow compensates the
change in $G$). As long as $G$ is constant there is no problem.
But if $G$ behaves similar to the coupling constants in
gauge-theories, it might be subject to a (moderate) rescaling at
high energies.\footnote{Note that Einstein's theory of general
relativity is difficult to quantize, if the "normal" quantization
procedure for a gauge-invariant theory on a fixed space-time
background are used. Only by exploiting the (additional) symmetry
of diffeomorphism invariance could a non-perturbative quantization
of gravity be achieved. Quite unexpectedly the (non-perturbative)
quantization turned out to be finite to all orders. It might well
be that the diffeomorphism-invariance of Einstein's theory of
gravitation provides a greater "protection" for the gravitational
(coupling) constant $G$ against renormalization than the gauge
symmetries in the typical gauge theories over a fixed background,
so that in fact $G$ might be constant at all energy scales.}

Assuming that $G$ might vary can regarded as a new type of
symmetry. Why should $G$ be considered as variable? Obviously $G$
has a definite value, at least at the typical energy scales
encountered today. On the other hand, in the purely classical
sector of general relativity, i.e. if we disregard quantum effects
(such as the discreteness of the geometry and the existence of a
fundamental area and fundamental particles), the gravitational
constant $G$ is just a convention which depends on the chosen
length or time scale. If macroscopic classical general relativity
is to be truly relational in the full Machian sense, i.e. only the
relative positions and sizes of "real macroscopic objects" count and no
fundamental reference scale (such as the minimum boundary area of
a fundamental quantum of geometry or particle) is given, the
macroscopic phenomena predicted by general relativity and
statistical thermodynamics of great numbers should be scale
invariant, i.e. exactly the same, whatever length scale we chose
and whatever the particular value of the gravitational constant
$G$ with respect to this length scale may be.

Therefore let us try to devise a reformulation of the results of
the previous sections so that all of the relevant phenomena, i.e.
the relations between entropy, local temperature, (Hawking)
temperature at infinity, metric etc. can be described in a way
that is essentially independent of $G$. Or formulated somewhat
differently: We are looking for an additional symmetry in the
equations which allows $G$ (or $\hbar G$) to vary without changing
the physical results predicted by the equations.

How can such a scale invariant reformulation be found? For this
purpose lets take a critical look on the Hawking entropy formula.
The Hawking entropy is equal to the surface area of a black hole,
measured (for example) in square meters, divided by four times the
Planck area, $4 \hbar G$. Via $G$ the Planck area can be expressed
in square meters as well, so that a dimensionless quantity arises
for the entropy. The problem is, that the area $A$ measured in
square meters, and the Planck area $\hbar G$ expressed in square
meters, might transform differently under a change of $G$: A shift
in $G$ will change the fundamental Planck-area, but according to
our common understanding of quantum theory such a shift shouldn't
affect our definition of the meter: The meter is defined via the
second which is linked to an atomic transition with a very sharply
defined frequency. The frequency of the atomic transition is a
pure quantum effect and therefore should not be affected by a
change in $G$.\footnote{This statement is only correct for quantum
theory on a fixed background. There are several objections to the
above statement, both practical and in principle: First, quantum
field theory on a fixed background is not diffeomorphism
invariant, and therefore should be regarded rather as an - albeit
excellent - approximation to the real physical phenomena. Second,
the transition frequency of an atomic clock depends on the value
of the fine-structure constant, which runs. However, this
dependence may be calculated in the context of QFT. Lastly there
is the practical problem to construct an atomic (or any other
accurate) clock at high energies.} Therefore the measured surface
area of a black hole or holostar (in square meters) should be
independent of $G$. On the other hand, the Planck area $\hbar G$
depends explicitly on $G$. If $G$ can vary, one would assume that
the divisor for the area in the entropy formula should rather be a
"fundamental area", whose measured value (for example in square
meters) is independent on a variation of $G$.

Is there an alternative to the divisor $4 \hbar G$ in the Hawking
formula? For the holostar solution we have the so called
fundamental length parameter, $r_0$, from which a fundamental area
$r_0^2$ can be formed. $r_0$ is defined as the maximum of the
radial metric coefficient $g_{rr}$ which attains its maximum at
the surface of the holostar, i.e. ${g_{rr}}_{max} = g_{rr}(r_h) =
r_h/r_0$. The area $r_0^2$ is related to the Planck area via
$r_0^2 = \beta \hbar G$. This is a significant improvement,
because of the factor $\beta$. We can compensate any change of the
gravitational constant $G$ by a variation of $\beta$. Furthermore,
the value of $\beta$ can be determined without reference to $G$,
as can be seen from equation (\ref{eq:T4:scaled:G}). A change in
$G$ therefore doesn't affect the measurement of $\beta$, which
seems just what is needed. Also note, that $\beta$ appears to be
just in the right range, i.e. $\beta = r_0^2/(\hbar G) \approx 4$,
so that the deviation to the Hawking formula wouldn't be large. In
the light of the above discussion we are lead to modify the
Hawking entropy formula as follows:

\begin{equation} \label{eq:S:scaled}
S = \frac{4 \pi r^2}{r_0^2} = \frac{1}{G} \, \frac{4 \pi
r^2}{(\beta \hbar)} = \left(\frac{G}{\beta \hbar}\right) 16 \pi
M^2
\end{equation}

$\beta = 4$ corresponds to the usual Hawking entropy-area law.

With this definition $F_E$ can be expressed in terms of $\beta$,
by setting the entropy of equation (\ref{eq:S:unfixed}) equal to
(\ref{eq:S:scaled}).

\begin{equation} \label{eq:FE:scaled}
\frac{F_E}{4 \pi \beta} = \left(\frac{8 \pi}{\beta}\right)^4
\end{equation}

The local temperature is given by equation (\ref{eq:TlocTherm})

\begin{equation} \label{eq:T4:scaled}
T^4 = \frac{1}{G} \, \frac{\left(\beta \hbar\right)^3}{2^{16}\pi^4
r^2}= \left(\frac{\beta \hbar}{G}\right)^3 \frac{1}{2^{18}\pi^4
M^2}
\end{equation}

$M$ is the gravitating mass of the holostar enclosed by the radius
$r$. The Hawking temperature reads:

\begin{equation} \label{eq:THawking:scaled}
T_{\infty} = \frac{(\beta \hbar)}{16 \pi r_h} = \left(\frac{\beta
\hbar}{G}\right) \frac{1}{32 \pi M}
\end{equation}

All the other equations of the previous section remain the same. A
very interesting thing has happened. With our particular ansatz
for the entropy (see equation (\ref{eq:S:scaled})), $\beta$ always
appears in combination with $1 / G$ (or rather $\hbar / G$) in all
equations, when the quantities are expressed in terms of the
gravitating mass $M$ (i.e. in "mass-coordinates"). On the other
hand, when the quantities are expressed in terms of the radial
coordinate $r$, i.e. in terms of length, $\beta$ always appears in
combination with $\hbar$.

Note also that equation (\ref{eq:T4:scaled}) relates the local
temperature to the local energy density $\rho = 1 /(8 \pi r^2 G)$.
If expression (\ref{eq:T4:scaled}) is expressed in terms of the
measurable quantities $\rho$ and $T$, the relation turns out to be
independent of $G$:

\begin{equation} \label{eq:T4:scaled:G}
T^4 = \frac{1}{G} \, \frac{\left(\beta \hbar\right)^3}{2^{16}\pi^4
r^2} = \frac{\left(\beta \hbar\right)^3}{2^{13}\pi^3} \rho
\end{equation}

In fact, all thermodynamic equations, such as $\partial S /
\partial E = 1/T$ are independent of $G$.

We can now analyze what happens if $G$ (or $\hbar$) vary. Let us
first assume $G = const$, but $\hbar$ variable. Any variation in
$\hbar$ can then be compensated by a respective change in $\beta$,
as long as $\hbar \beta$ remains constant. I.e $\beta$ must change
inverse proportional to $\hbar$. On the other hand, if $G$ varies
and $\hbar$ is constant, a change in $G$ will not affect the
equations (expressed in "mass-coordinates"), whenever $\beta$
varies proportional to $G$, as $G$ always appears in combination
with $\beta \hbar/G$.

Again there is some sort of duality between the quantities $G$ and
$\hbar$.

We are confronted with two different choices for the
Hawking-entropy formula. Either we stick to the classical result
$S = \frac{A}{4 \hbar}$ or we chose equation (\ref{eq:S:scaled}).
The equations for the internal temperature within the holostar,
the (Hawking) temperature at infinity and the number of degrees of
freedom $F_E$ will be affected by the choice, as can be seen from
equations (\ref{eq:THawking:scaled}, \ref{eq:T4:scaled},
\ref{eq:FE:scaled}).

What particular choice we should make, depends on the nature of
the gravitational constant $G$. If $G$ is a universal constant of
nature, independent of the energy scale, i.e. not subject to a
(moderate) renormalization at high energies, we should use the
original Hawking entropy formula. However, if $G$ undergoes
renormalization at high energies, it appears more appropriate to
pick the more general approach with $\beta$ in the divisor of the
entropy formula.

From an aesthetic point of view, the second choice has some merit.
However, not aesthetics, but measurements will have to decide the
issue. In the previous section it has been demonstrated, that
under the assumption that $G$ is constant, the Hawking-entropy
formula is correct to better than 1\%. If the entropy formula is
to be modified according to the discussion of this section, the
value of $\beta$ can be determined experimentally along the same
lines as in the previous section. Equation (\ref{eq:T4:scaled:G})
gives a relation between the local radiation temperature and the
total local energy density within the holostar:

\begin{equation} \label{eq:T4:scaled:rho}
T^4 = \left(\frac{\beta \hbar}{8 \pi}\right)^3 \frac{\rho}{16}
\end{equation}

This can be solved for $\beta$:

\begin{equation} \label{eq:beta:from:T4:rho}
\beta = \left(\frac{(2T)^4}{\rho} \right)^{\frac{1}{3}} \frac{8
\pi}{\hbar}
\end{equation}

With $T_{CMBR} = 2.725 K$ and $\rho_{WMAP} = 2.465 \cdot 10^{-27}
\frac{kg}{m^3}$ we find:

\begin{equation}
\beta = 4.17
\end{equation}

When this value is compared with equation (\ref{eq:beta:running}),
it turns out far too high for the low temperature region of the
universe, where the fine-structure constant $\alpha$ is small.
Furthermore, it doesn't fit well into the framework developed in
\cite{petri/charge, petri/hol}. Therefore at the current state of
knowledge it appears more likely, that the Hawking entropy formula
is correct for all scales, which in turn can be interpreted as
indication, that the gravitational constant $G$ is a true constant
of nature such as $\hbar$ and $c$, i.e. independent of the energy
scale.

On the other hand, if the total matter density $\rho$ would turn
out to be roughly $13 \%$ higher than the WMAP result, we would
get $\beta = 4$, which would make both approaches essentially
indistinguishable.

\section{Are the gravitational constant, the fine-structure constant and the electron mass related?}

In this section I propose a relation between the gravitational
constant $G$, the fine-structure constant $\alpha$ and the
electron-mass $m_e$. This proposal most likely will be considered
by any conservative researcher as a blind leap into the dark,
rather fuelled by the faith of the foolish than by the wisdom of
knowledge. And not much can be said against such a point of view:
The relationship proposed in equation (\ref{eq:G}) is based more
on "playing with numbers" than on truly convincing physical
arguments. Therefore it is quite probable that (\ref{eq:G}) will
suffer the fate of the vast majority of proposals of a similar
kind, such as the numerous formula given for the value of the
fine-structure constant, which even appear in regular intervals
today.\footnote{It requires quite a bit of faith to attribute a
definite numerical value to a running quantity.} On the other
hand, the gravitational constant is the only fundamental constant
of nature that hasn't been measured to a high degree of precision.
Therefore from the viewpoint of metrology it would be helpful, if
$G$ could be related to quantities that have been measured
precisely. Furthermore, the discussion in the previous section
might be viewed as some - albeit very tentative - evidence, that
$G$ and the fundamental quantities that determine our length and
time-scales, such as $\alpha$ and $m_e$, should be related in one
way or the other: In the previous section it has been remarked,
that at high energies our definition of the meter, which is linked
to the second via the speed of light, $c$, might change due to the
running of the coupling constant(s). A change in the length scale
will affect the maximum entropy of a space-time region bounded by
a proper area $A$, unless there is a definite relation between the
gravitational constant $G$ and the fundamental quantities or
processes that define our length or time scale at any particular
energy, which allows us to calculate the change.

How can we define a time- or length scale at an arbitrary energy?
In order to do this we need a physically realizable clock, which
works (at least in principle) over the whole energy range from the
Planck-energy to the low energies we encounter today. In order to
define a time-interval as reference, we need a massive particle
(time has no meaning for photons). This particle should be
available throughout the whole energy range. From a practical
point of view it should also be fundamental, meaning not
composite, so that its properties and behavior can be "easily"
calculated. The only particle which fulfils these conditions is
the electron. At low energies we could construct a positronium
clock. At high energies it might be more appropriate to use the
compton wavelength of the electron as reference. In any case, the
fundamental time or length interval determined from our electron
based clock will depend on the energy scale, due to the running
value of the fine-structure constant and the running electron
mass. Unless there is a definite relationship between $G$,
$\hbar$, $\alpha$ and $m_e$ it will be difficult to determine the
entropy of a self gravitating system in an unambiguous way.

How could such a relationship look like? The quantity $\hbar c /
G$ has the dimension of mass-squared. In order to get the
dimension right, we should start with

\begin{equation}
\hbar c / G = m_{Pl}^2 \propto m_e^2
\end{equation}

Unfortunately there is a great discrepancy between the Planck-mass
and the electron-mass. Nevertheless, $m_{Pl}/m_e$ is a
dimensionless quantity, that might be related to the only other
dimensionless fundamental quantity available at the low-energy
scale, $\alpha$. Therefore we make the following ansatz:

\begin{equation} \label{eq:ln:mpl:me}
\ln{\left(\kappa \frac{m_{Pl}}{m_{e}}\right)} = \frac{x}{ \alpha}
\end{equation}

This ansatz is motivated by the renormalization group equations,
according to which the coupling constants vary with energy with

\begin{equation}
\frac{1}{\alpha(E)} - \frac{1}{\alpha(E_0)} = \frac{b}{2 \pi}
\ln{\frac{E}{E_0}}
\end{equation}

where $b$ depends on the model.\footnote{$b[U(1)] = -\frac{4}{3}
N_g$ and $b[SU(N)]= \frac{11}{3} N - \frac{4}{3} N_g$, where $N_g$
is the number of generations (three in the standard model).}

What could be a reasonable value for $x$? There is some "evidence"
for $x = 3/8$. In the simplest SU(5) GUT-theory the normalization
of the electric charge operator with respect to the other
operators requires this value. Furthermore, $3/8$ is the
prediction for the Weinberg angle at the GUT-energy in minimal
SU(5), i.e. the ratio of the electromagnetic coupling to the
"true" unified coupling constant at the unification energy.

With $x = 3/8$ and setting $\kappa$ temporarily to 1 equation
(\ref{eq:ln:mpl:me}) reads:

\begin{equation} \label{eq:ln:mpl:me:2}
\ln{\left(\frac{m_{Pl}}{m_{e}}\right)} \approx \frac{3}{8 \alpha}
\end{equation}

If we plug in the experimentally determined values for the
electron-mass, the fine-structure constant and the Planck-mass
(which requires knowledge of $G$, $\hbar$ and $c$), we find a not
too good agreement:

\begin{equation} \label{eq:ln:mpl:me:numerical}
\ln{\left(\frac{m_{Pl}}{m_{e}}\right)} = 51.5279
\end{equation}

and

\begin{equation} \label{eq:3/8a:numerical}
\frac{3}{8 \alpha} = 51.3885
\end{equation}

However, there is still one free parameter in the ansatz of
equation (\ref{eq:ln:mpl:me}). Furthermore, the motivation for the
search of a relation between $G$, $\hbar$, $m_e$ and $\alpha$ was
the observation, that the length-scale (or area scale) might
change at high energies. Therefore the fundamental area $r_0^2$,
which somehow "documents" this change, should enter into the above
relation. For an electrically charged spin 1/2 particle, such as
the electron, the radius $r_h$ of its boundary (or membrane) can
be found in \cite{petri/charge}:

\begin{equation}
\frac{r_h^2}{\hbar G} = \frac{\beta}{4} = \frac{\alpha}{2} +
\sqrt{\left(\frac{\alpha}{2}\right)^2 + \frac{3}{4}}
\end{equation}

Quite curiously, if we replace $m_e \rightarrow m_e / (\beta/4)$
the logarithms in equations (\ref{eq:ln:mpl:me:numerical},
\ref{eq:3/8a:numerical}) are equal within the measurement
errors\footnote{The error is dominated by the uncertainty of
$m_{Pl}$ due to the large uncertainty of $G$}:

\begin{equation}
\ln{\left(\frac{\beta}{4} \frac{m_{Pl}}{m_{e}}\right)} = 51.3883
\approx \frac{3}{8 \alpha} = 51.3885
\end{equation}

With $m_{Pl}^2 = \hbar c / G$ we can solve the above equation for
$G$:

\begin{equation} \label{eq:G}
G = \frac{\hbar c}{m_e^2}  \left(\frac{\alpha}{2} +
\sqrt{\left(\frac{\alpha}{2}\right)^2 + \frac{3}{4}}\right)^2
 e^{-\frac{3}{4 \alpha}}
\end{equation}

which gives the following "prediction" for $G$:

\begin{equation}
G = 6.670460 \cdot 10^{-11} \frac{m^3}{kg \, s^2}
\end{equation}

This is well within the errors of the value recommended by CODATA
1998 ($G =6.673(10) \cdot 10^{-11} m^3/(kg \, s^2)$ and within $3
\sigma$ of the "old" value $G = 6.6726(7) \cdot 10^{-11} m^3/(kg
\, s^2)$, recommended by the International Council of Scientific
Units 1986, which however was discarded by CODATA in 1998 because
of difficulties to reproduce this result.

Keep in mind, however, that this section is pure numerology, which
has not much to do with predictive science. As far as is known to
the author, there is no example in the history of science, where
numerology has guided us to any significant scientific result.

\section{\label{sec:Tables}Fermionic weighting factors for the energy- and number-density}

It is useful to cast the equations of the previous sections into a
more familiar form. Usually the effective degrees of freedom of a
gas of ultra-relativistic bosons and fermions are calculated by
weighting the fermion degrees of freedom with a factor $w = 7/8$.
This procedure allows us to apply the familiar Planck formula for
the energy-density of a photon gas to the arbitrary case of a gas
of ultra-relativistic bosons and fermions, simply by replacing the
two photon degrees of freedom\footnote{Note, that in our
convention for the counting of the degrees of freedom $g=2$
corresponds to $f_B=1$} in the Planck-formula with the effective
degrees of freedom of the fermions and bosons.

The factor $7/8$ for each fermionic degree of freedom is nothing
else than the ratio of $Z_{F,3} (u_F = 0)$ by $Z_{B,3}(u_B = 0)$.
This ratio is relevant for the determination of the
energy-density.\footnote{The ratio $Z_{F,2}(0)/Z_{B,2}(0) = 3/4$
is relevant for the number-density} The weighting factor $7/8$ for
the fermionic degrees of freedom is only correct, when $u_F = u_B
= 0$, i.e. when the chemical potential of all particles is zero.
For the holostar in thermodynamic equilibrium $u_F = 0$ is not
possible, at least not in the simple ultra-relativistic model
discussed here. $u_F$ is always larger than $1.344$ and depends on
the ratio of the (unweighted) degrees of freedom of the
ultra-relativistic bosons and fermions. Nonetheless, we can still
use the standard Planck-formula for a photon gas by adhering to
the following procedure:

\begin{itemize}
\item Determine the bosonic and fermionic particle degrees of freedom,
$f_B$ and $f_F$. \item Calculate $u_F$ as a function of $f_B/f_F$
from equation (\ref{eq:ImplicituF}). \item Determine the fermionic
weighting factors $w_E$ and $w_N$ for matter and anti-matter
respectively. The weights only depend on $u_F$.  \item Use $w_E$
instead of $7/8$ and $w_N$ instead of $3/4$ as the appropriate
weighting factors for the fermionic degrees of freedom in order to
determine the effective degrees of freedom of the whole gas.
\end{itemize}

If we denote by $\widetilde{f_E}$ the effective degrees of freedom
required for the determination of the energy density and by
$\widetilde{f_N}$ the effective degrees of freedom required for
the number-density we get:

\begin{equation}
\widetilde{f_E} = 2 f_B + (w_E + \overline{w_E}) f_F
\end{equation}

\begin{equation}
\widetilde{f_N} = 2 f_B + (w_N + \overline{w_N})f_F
\end{equation}

with

\begin{equation} \label{eq:wE}
w_E = \frac{Z_{F,3}(u_F)}{Z_{B,3}(0)}
\end{equation}

and

\begin{equation} \label{eq:wN}
w_N = \frac{Z_{F,2}(u_F)}{Z_{B,2}(0)}
\end{equation}

The weighting factors for the fermionic anti-matter $\overline{w_E}$ and
$\overline{w_N}$ are given by the respective negative value of the
fermionic chemical potential per temperature, i.e. $u_F \rightarrow -u_F$:

\begin{equation}
\overline{w_E} = w_E(-u_F) = \frac{Z_{F,3}(-u_F)}{Z_{B,3}(0)}
\end{equation}

and

\begin{equation}
\overline{w_N} = w_N(-u_F) =\frac{Z_{F,2}(-u_F)}{Z_{B,2}(0)}
\end{equation}

The quantity $w_E$ gives the ratio of the total energy density  of
a single fermionic degree of freedom to a single bosonic degree of
freedom (with zero chemical potential) in an arbitrary volume, i.e.

\begin{equation}
w_E = \frac{\delta E_F}{\delta E_B}
\end{equation}

whereas $w_N$ gives the ratio of the total number of fermions to
bosons (for a single bosonic and fermionic degree of freedom) in
an arbitrary volume:

\begin{equation}
w_N = \frac{\delta N_F}{\delta N_B}
\end{equation}

The respective ratios of the energy- and number-densities of any
fermion with respect to its anti-particle are then given by $w_E /
\overline{w_E}$ and $w_N / \overline{w_N}$

The mean energy per particle is proportional to the temperature.
However, the constant of proportionality is different for fermions
and bosons:

\begin{equation}
\frac{\delta E_B}{\delta N_B} = \varepsilon_B T
\end{equation}

and

\begin{equation}
\frac{\delta E_F}{\delta N_F} = \varepsilon_F T
\end{equation}

with

\begin{equation} \label{eq:eB}
\varepsilon_B =\frac{\widetilde{f_N}}{\widetilde{f_E}} \, \sigma
\end{equation}

but

\begin{equation} \label{eq:eF}
\varepsilon_F = \frac{w_E}{w_N}
\frac{\widetilde{f_N}}{\widetilde{f_E}} \, \sigma
\end{equation}

The ratio of the mean energy of a single fermionic degree of
freedom to a single bosonic degree of freedom is given by:

\begin{equation}
\frac{\varepsilon_F}{\varepsilon_B} = \frac{w_E}{w_N}
\end{equation}

The above dependencies guarantee, that the mean energy per
particle per temperature $\varepsilon$ (for all particles, i.e.
including all fermions and bosons) is equal to the mean entropy
per particle $\sigma$ (again for all particles):

\begin{equation}
\frac{\delta E}{\delta N} = \varepsilon T = \sigma T
\end{equation}

with

\begin{equation}
\varepsilon = \frac{f_F (N_F \varepsilon_F + \overline{N_F}
\overline{\varepsilon_F})+ 2 f_B N_B \varepsilon_B}{f_F (N_F +
\overline{N_F})+ 2 f_B N_B} = \varepsilon_B
\frac{\widetilde{f_E}}{\widetilde{f_N}} = \sigma
\end{equation}

Note that the relation $\varepsilon = \sigma$, which can be seen
as the basic thermodynamical characteristic of the holostar, is
only true for the mean energy and entropy of all particles, but
not for the individual particle species, for which the relations
(\ref{eq:eB}, \ref{eq:eF}) hold. These relations in general imply
$\varepsilon_B \neq \sigma_B$ and $\varepsilon_F \neq \sigma_F$,
unless the number of bosonic degrees of freedom is zero.

The mean entropy per particle $\sigma$ can be calculated from the
mean entropy per fermion and boson:

\begin{equation}
\sigma = \frac{f_F (\sigma_F w_N + \overline{\sigma_F}
\overline{w_N}) + 2 f_B \sigma_B}{\widetilde{f_N}}
\end{equation}

With all of the above definitions equation (\ref{eq:FE:beta})
reads as follows\footnote{In the scale-invariant reformulation of
the Hawking entropy in section \ref{sec:Hawking:scaled} we have to
use equation (\ref{eq:FE:scaled}), which then is given by $ 4 \pi
\beta = \frac{\widetilde{f_E} }{15 \cdot 16}
\left(\frac{\beta}{4}\right)^4$}:

\begin{equation} \label{eq:beta:fw}
4 \pi \beta = \frac{\widetilde{f_E} }{15 \cdot 16}
\end{equation}

The effective degrees of freedom determined by the above procedure
can be plugged into the well known equations for the
energy-density, the entropy density and the number-density of a
photon gas. For example, the energy-density is now given as:

\begin{equation} \label{eq:rho:fe}
\rho = \frac{1}{8 \pi r^2} = \widetilde{f_E} \frac{\pi^2}{30
\hbar^3} T^4
\end{equation}

In Table \ref{tab:thermo} the relevant thermodynamic quantities
are compiled as a function of the ratio $f_B/f_F$ of bosonic to
fermionic degrees of freedom. All the quantities in the table are
normalized to the thermodynamic quantities of an ideal
relativistic boson gas. Recall, that for a relativistic boson gas
$\sigma_B = 2 \pi^4 / (45 \zeta(3)) = 3.60157$ and $\varepsilon_B
= \pi^4 / (30 \zeta(3)) = 2.70118$.

\begin{table}
\begin{center}

\begin{tabular} {c||c|c|c|c|c|c|c|c}
$f_B/f_F$ & $u_F$ & $\sigma$ & $\sigma_F$ &
${\overline{\sigma_F}}$ & $w_N$ & $\overline{w_N}$ & $w_E$ &
$\overline{w_E}$ \\ \hline \hline
0    & 1.34416 & 3.3792 & 3.2006 & 5.4052 & 2.3865 & 0.2103 & 3.0114 & 0.2372 \\
1/100& 1.35046 & 3.3758 & 3.1966 & 5.4112 & 2.3986 & 0.2091 & 3.0282 & 0.2357 \\
1/10 & 1.40496 & 3.3489 & 3.1615 & 5.4626 & 2.5054 & 0.1983 & 3.1766 & 0.2234 \\
1/5  & 1.46148 & 3.3248 & 3.1256 & 5.5162 & 2.6204 & 0.1877 & 3.3375 & 0.2113 \\
1/3  & 1.53124 & 3.2993 & 3.0820 & 5.5824 & 2.7685 & 0.1753 & 3.5462 & 0.1972 \\
1/2  & 1.61097 & 3.2750 & 3.0330 & 5.6584 & 2.9464 & 0.1622 & 3.7991 & 0.1823 \\
1    & 1.81380 & 3.2299 & 2.9126 & 5.8529 & 3.4424 & 0.1330 & 4.5175 & 0.1492 \\
2    & 2.12023 & 3.1902 & 2.7421 & 6.1493 & 4.3213 & 0.0984 & 5.8340 & 0.1101 \\
3    & 2.35325 & 3.1735 & 2.6213 & 6.3764 & 5.1053 & 0.0782 & 7.0515 & 0.0873 \\
5    & 2.70549 & 3.1607 & 2.4524 & 6.7219 & 6.5032 & 0.0552 & 9.3134 & 0.0615 \\
8    & 3.08811 & 3.1568 & 2.2865 & 7.0994 & 8.3468 & 0.0377 & 12.456 & 0.0420 \\
10   & 3.28970 & 3.1575 & 2.2059 & 7.2989 & 9.4688 & 0.0309 & 14.448 & 0.0344 \\
100  & 6.21570 & 3.2236 & 1.4151 & 10.216 & 41.804 & 0.0017 & 88.571 & 0.0019 \\
1000 & 11.3315 & 3.3308 & 0.8378 & 15.332 & 217.25 & $10^{-5}$ & 734.05 & $10^{-5}$ \\
3000 & 14.9899 & 3.3793 & 0.6436 & 18.990 & 487.51 & $3 \cdot 10^{-7}$ & 2116.2 & $3 \cdot 10^{-7}$ \\
\end{tabular}
\caption{\label{tab:thermo}Thermodynamic parameters for an
ultra-relativistic gas of fermions and bosons, compiled for
selected ratios of bosonic to fermionic degrees of freedom
$f_B/f_F$. $u_F$ is the dimensionless chemical potential per
temperature of the fermions. Anti-fermions have the opposite
value. The chemical potential of the bosons is zero. $\sigma$ is
the mean entropy per particle, $\sigma_F$ is the entropy per
fermion. $w_E$ and $w_N$ are the weighting factors for the energy-
and number-densities with respect to a bosonic degree of freedom.
Barred quantities refer to anti-fermions.}

\end{center}
\end{table}

We find, that the fermions within the holostar have much higher
weights for the energy- and number-densities than the ordinary
weights of $7/8$ or $3/4$, whereas the anti-fermions have much
lower weights. For example, when $f_F = f_B$, the weighting factor
for the fermionic energy-density, $7/8$, must be replaced by $4.5$
and that for the anti-fermion with $0.15$.

An interesting case is $f_B = 8 f_F$, i.e. when there are 8 bosons
per fermion. For this case the entropy per particle $\sigma$ (and
the energy per particle) within the holostar is minimized. Note
that $\sigma = 3.1568$ is only slightly larger than $\pi$: $\sigma
/ \pi = 1.00485$.

There is a curious modification to the thermodynamic model, which
might be of some interest. If we formally assume, that the
fermions have no anti-particles, we can give the bosons a negative
chemical potential exactly opposite to that of the fermions,
$\mu_F + \mu_B = 0$, and still retain a symmetric description. In
a formal sense the bosonic partner particle of any fermion,
carrying the opposite chemical potential, might be considered as
the anti-particle of its fermionic counterpart. Let's call this
peculiar matter phase the "abnormal supersymmetric phase". We can
construct a table similar to Table \ref{tab:thermo}. The
respective weighting factors for the bosonic energy- and
number-densities are given by:

\begin{equation}
{w_E}_B = \frac{Z_{B,3}(-u_F)}{Z_{B,3}(0)}
\end{equation}
\\

\begin{equation}
{w_N}_B = \frac{Z_{B,2}(-u_F)}{Z_{B,2}(0)}
\end{equation}
\\

The essential thermodynamic parameters of the "abnormal"
supersymmetric phase are compiled in Table \ref{tab:super}

\begin{table}
\begin{center}

\begin{tabular} {c||c|c|c|c|c|c|c|c}
$f_B/f_F$ & $u_F$ & $\sigma$ & $\sigma_F$ & $\sigma_B$ & $w_E$ & $w_N$ & ${w_E}_B$ & ${w_N}_B$ \\
\hline \hline
0    & 1.11721 & 3.3516 & 3.3516 & 5.0269 & 2.4620 & 1.9842 & 0.3089 & 0.2846 \\
1/10 & 1.14848 & 3.3534 & 3.3304 & 5.0613 & 2.5318 & 2.0359 & 0.2992 & 0.2754 \\
1/3  & 1.21268 & 3.3581 & 3.2872 & 5.1314 & 2.6808 & 2.1457 & 0.2802 & 0.2575 \\
1    & 1.35321 & 3.3717 & 3.1948 & 5.2835 & 3.0355 & 2.4039 & 0.2428 & 0.2225 \\
3    & 1.61169 & 3.4059 & 3.0325 & 5.5589 & 3.8015 & 2.9480 & 0.1868 & 0.1704 \\
10   & 2.03254 & 3.4787 & 2.7895 & 5.9986 & 5.4263 & 4.0529 & 0.1221 & 0.1108 \\
100  & 3.12422 & 3.7293 & 2.2717 & 7.1131 & 12.795 & 8.5398 & 0.0407 & 0.0368 \\
1000 & 4.45250 & 4.1130 & 1.8183 & 8.4496 & 31.935 & 18.341 & 0.0108 & 0.0097 \\
\end{tabular}
\linebreak \caption{\label{tab:super}Thermodynamic parameters for
the "abnormal supersymmetric phase" of an ultra-relativistic gas
consisting of fermions and bosons, compiled for selected ratios of
the bosonic to fermionic degrees of freedom $f_B/f_F$. $u_F$ is
the chemical potential per temperature of the fermions, the bosons
have the opposite value. $s$ is the mean entropy per particle,
$\sigma_F$ and $\sigma_B$ are the entropies per fermion and boson,
respectively. $w_E$ and $w_N$ are the weighting factors for the
energy- and number-densities of the fermions. The weighting
factors of the bosons are denoted by subscript $B$.}
\end{center}
\end{table}

It is a curious numerical coincidence, that for the "abnormal
supersymmetric phase" with identical bosonic and fermionic
particle degrees of freedom the ratio of the entropy per boson
$\sigma_B$ to the entropy per fermion $\sigma_F$ in the holostar
is almost equal to the ratio predicted for the respective areas of
a single spin-1 spin-network state (=boson?) to a single spin 1/2
spin-network state (=fermion?) in loop quantum gravity (LQG):

\begin{equation}
\left(\frac{\sigma_B}{\sigma_F}\right)_{holo} \simeq 1.654
\end{equation}

whereas:

\begin{equation}
\left(\frac{\sigma_B}{\sigma_F}\right)_{LQG} =
\sqrt{\frac{j_B(j_B+1)}{j_F(j_F+1)}} = \sqrt{\frac{8}{3}} \simeq
1.633
\end{equation}

Whether this finding is significant, is hard to tell. Although
there appears to be a connection between the ultra-relativistic
particles of the holostar solution with the links of a loop
quantum gravity (LQG) spin-network state, it is yet too early to
draw any definite conclusions. See \cite{petri/charge} for a more
detailed discussion on the possible connection between LQG and the
classical holographic solution.

There is another interesting observation. If we compare the first
line in Table \ref{tab:thermo}, i.e. the case of a "normal" gas
consisting exclusively out of ultra-relativistic fermions and
anti-fermions, with the $f_F = f_B$ line in Table \ref{tab:super},
i.e. the "abnormal supersymmetric-phase" with equal fermionic and
bosonic degrees of freedom, we find that the thermodynamic
properties of both phases are very similar. The chemical
potentials per temperature are $u = 1.344$ in the first case and
$u = 1.353$ in the second case. The mean entropy per particle is
$\sigma = 3.379$ in the first case and $\sigma = 3.372$ in the
second case. If we compare the other quantities in the table, i.e.
the entropies of the fermions, the entropies of the
anti-fermions/bosons, the weighting factors for the fermions, the
weighting factors for the anti-fermions/bosons, we also find, that
all of these quantities are very similar. Therefore, from a purely
thermodynamic point of view, a "normal" ultra-relativistic
gas-phase consisting only out fermions and anti-fermions has
nearly identical properties to the "abnormal supersymmetric"
gas-phase consisting out of an equal number of fermions and
bosons, with the interpretation that the bosons have "disguised"
themselves as the anti-particles of the fermions. So in a strictly
formal sense one could say, that at ultra-high temperatures a gas
consisting exclusively out of fermions and their anti-particles
becomes more or less indistinguishable from a gas consisting out
of equal numbers of fermions and bosons.

\section{\label{sec:asymmetry}On the matter-antimatter asymmetry in curved space-times}

In the previous sections we have seen, that there is only a
solution to the thermodynamic constraint equation $F_S(u_F) =
F_E(u_F)$, when we have at least one ultra-relativistic fermionic
species present. No matter what the specifics of the thermodynamic
model are, the ultra-relativistic fermions are required to have a
non-zero chemical potential, which is significantly higher than
the local radiation temperature. We can loosely interpret this
finding such, that we need the degeneracy pressure of at least one
ultra-relativistic fermion in order to stabilize the self
gravitating object, so that it doesn't collapse under its own
gravity to a singularity.

Table \ref{tab:thermo} describes the characteristic properties of
an ideal gas of ultra-relativistic fermions and bosons, which are
in thermal equilibrium with each other and their anti-particles.
The non-zero chemical potential of the fermions induces an
asymmetry in the relative number-densities of a fermionic particle
and its anti-particle. From this asymmetry different values for
the entropy per particle/anti-particle and the ratio of the
energy-densities and number-densities arise.

The asymmetry is smallest, when at a certain spatial position
within the holostar there are only fermions and no bosons. For
this situation the chemical potential per temperature $u_F$
attains its minimum value of $u_F \simeq 1.34416$.

If we increase the number of bosonic species with respect to the
fermions, the matter-antimatter asymmetry in the fermions becomes
higher with increasing $u_F$, as can be seen from comparing the
first and last columns of the Table \ref{tab:thermo}. For $f_B =0$
we have $w_E \simeq 3.01$ and $\overline{w_E} \simeq 0.24$, so
that the ratio of the energy-densities of matter vs. anti-matter
in any proper volume (where the fermions still are
ultra-relativistic) is: $\eta_E = 12.6967$. The number-densities
have roughly the same ratio: $\eta_N = \overline{w_N} / w_N =
11.3453$. When the number of fermionic and bosonic species is
equal, we find that $\eta_E = 30.2894$ and $\eta_N = 25.8847$,
i.e. a significantly higher asymmetry.

The holsostar's interior structure, or rather the condition
$\sigma = \epsilon / T$, induces a natural asymmetry between the
fermionic matter and antimatter in thermodynamic equilibrium in
the spatial holostar metric. Only the fermions are effected in
such a way. For the bosons there is no such shift, because they
cannot have a non-zero chemical potential, at least as long as
they are ultra-relativistic.

If the interior temperature falls below the mass-threshold of a
particular fermionic species, fermions and anti-fermions will
annihilate. Due to the large asymmetry above the threshold most
fermions will survive the annihilation process. The ratio of the
number of surviving fermions with respect to the total number of
fermions and anti-fermions above the threshold is:

$$ \eta = \frac{\eta_N - \overline{\eta_N}}{\eta_N + \overline{\eta_N}} $$

This ratio attains its minimum value in the case where there are
only fermions, no bosons, i.e. $f_B/f_F = 0$, where $\eta \approx
84 \%$. The ratio can be significantly higher, if the number of
bosonic species is higher.

In \cite{petri/hol} it has been shown, that the holostar has some
potential to serve as an alternative model for the universe. If
this truly turns out to be the case, the thermodynamic properties
of the holostar solution naturally explain the matter-antimatter
asymmetry in the universe: When during the expansion the
temperature falls below the rest-mass of a particular fermionic
species, for example the baryons (or quarks, electrons), the
fermions will annihilate with their anti-partners. Due to the
curvature-induced matter-antimatter asymmetry there will be at
least a factor of 11 more particles than antiparticles, so that
the mutual annihilation conserves most of the energy-density
within any particular fermionic species, at least 84 \%. This
figure might be significantly higher, when there are many more
bosonic than fermionic species. However, in the Standard Model of
particle physics the situation is rather the other way around: We
have more fermions than bosons. Whenever $f_F \geq f_B$, i.e. the
number of fermionic degrees of freedom is larger or equal than the
bosonic degrees of freedom, the percentage of fermions surviving
the annihilation process with respect to the total number of
fermions before the annihilation started, is only very moderately
dependent on the ratio $f_B / f_F$. It ranges from $84 \%$ at $f_B
= 0$ over $88 \%$ at $f_F = 3 f_B$ to $92.5 \%$ at $f_F = f_B$.

This effect also explains, why $r_0^2$, which has been shown to
depend linearly on the effective ultra-relativistic degrees of
freedom, has a nearly universal value, although the number of
relativistic particle degrees of freedom is dramatically reduced
during the expansion: Due to the large matter-antimatter asymmetry
the energy-density in a particular particle species is nearly
conserved, when the species "freezes out", so that the effective
number of the degrees of freedom at the transition doesn't change
significantly.

This observation is significant in two respects:

First it allows us to interpret the energy density in the universe
as we find it today, as a fairly good indicator of the number of
degrees of freedom at very high energies, where all particle
species are expected to be ultra-relativistic. In fact, the
"effective" number of degrees of freedom, determined via the
energy-density of the matter today, is expected just to slightly
underestimate the total number of degrees of freedom at the Planck
scale. In the worst case the observed total energy density would
be roughly 85\% of the energy-density at a temperature, where all
of the particles were relativistic. We will see in the next
section, that this appears to be actually the case.

Second, it gives a good a-posteriori justification for our very
early assumption, that the fundamental area $r_0^2$ is a (nearly)
universal quantity, constant whenever the universe doesn't undergo
a phase-transition, and which "runs" only moderately with the
energy-scale via the coupling-constant(s), whose values are
related to the particle degrees of freedom that are available at a
given energy.

Whereas the interpretation of $r_0^2$ as a running area scale,
only slightly dependent on the energy-scale, can be considered to
be backed by both observational and theoretical insight, the
particular form of $r_0^2$ as proposed in equation
(\ref{eq:beta:running}) has not such a sound justification, at
least at high energies. Our understanding of the holostar solution
and particle physics in curved space-times will improve
dramatically, if the apparent correspondence between the
fundamental length- and energy-scales, the number of particle
degrees of freedom and the running of the coupling constant(s) can
be made more definite.

\section{An estimate for the number of degrees of freedom at the Planck scale}

With the discussion beforehand and the tables given in section
\ref{sec:Tables} one can estimate the number of degrees of freedom
at the Planck energy, where all particle degrees of freedom are
expected to be ultra-relativistic.

I will only discuss the "normal" matter phase, whose properties
are given by Table \ref{tab:thermo}. With $f$ let us denote the
total number of degrees of freedom, i.e. $f = 2(f_B+f_F)$. If
supersymmetry is a true symmetry of nature at high energies, we
should expect the bosonic and fermionic degrees of freedom to be
equal at the Planck energy, which gives: $f_B = f_F = f/4$.

From equation (\ref{eq:FE:beta}) we find:

\begin{equation} \label{eq:fE:Planck}
\widetilde{f_E} = \frac{f}{2}(1+\frac{w_E+\overline{w_E}}{2}) = 2^6 \cdot 3 \cdot 5 \cdot
\pi \cdot \beta
\end{equation}

and therefore\footnote{If the Hawking entropy area law were modified according to the discussion in section \ref{sec:Hawking:scaled}, the above
expression must be multiplied with $(4/\beta)^4$:
$$f =  \frac{2^7 \cdot 3 \cdot 5 \cdot \pi \cdot \beta}{(1+\frac{w_E+\overline{w_E}}{2})}
\left(\frac{4}{\beta}\right)^4 $$
}:

\begin{equation} \label{eq:f:Planck}
f = \frac{2^7 \cdot 3 \cdot 5 \cdot \pi \cdot \beta}{(1+\frac{w_E + \overline{w_E}}{2})}
\end{equation}

For $f_F = f_B$ one can show (see \cite{petri/asym}):

\begin{equation} \label{eq:wewe}
\frac{w_E + \overline{w_E}}{2} = \frac{7}{3}
\end{equation}

so that:

\begin{equation} \label{eq:f:Planck:super}
f =  2^4 \cdot 3^2 \cdot 4 \pi \beta
\end{equation}

In order to determine $f$ we need to know $\beta = r_0^2 / \hbar$
at the Planck scale. In \cite{petri/charge} arguments were given,
that $\beta/4 \approx \sigma / \pi \simeq 1$ at the Planck
energy\footnote{It is suggestive to set $\beta = 4$ at the
Planck-energy, which at the same time will set $\alpha = 1/4$.
This would make the prediction $f = 2^6 3^2 4 \pi$. The factor of
$\pi$ is somewhat disappointing. One would expect an integer
number, at least if particles were the truly fundamental building
blocks of nature. The basic building blocks of the holostar,
however, appear to be rather strings and membranes. Therefore it
is conceivable, that the particle degrees of freedom must be
regarded as an effective description. For an effective description
non-integer values for $f$ are not uncommon. The full number of
degrees of freedom at or above the Planck-scale, including the
"stringy" degrees of freedom, is expected to be higher and
integer. Only a unified theory of quantum gravity, which most
likely will be based on string-theory, will be able to tell us,
what the full spectrum of the basic building blocks of nature,
i.e, their interior structure and their relative abundances, is
going to be.}. A more conservative estimate will place $\beta$ in
the following range:

\begin{equation} \label{eq:range:beta}
\sqrt{\frac{3}{4}} \leq \frac{\beta}{4} \leq \frac{\sigma}{\pi}
\end{equation}

For the above range we find from equation (\ref{eq:f:Planck:super}):

\begin{equation} \label{eq:f:Planck:2}
6269 < f < 7442
\end{equation}

Note that the ratio of the lower to the higher number in this
range is very close to of 84 \%. This is almost exactly the ratio
one expects for the effective degrees of freedom at low vs. high
energies due to the matter-antimatter asymmetry (see section
\ref{sec:asymmetry}). In fact, $6269 / 7442 = 0.842$.

For $\beta = 4$ we have $f \approx 7238$. These are all quite
large numbers compared to the number of particle degrees of
freedom of the Standard Model. On the other hand it has been
speculated, that at exceedingly higher energies more and more new
particles will show up, and that this process might continue
indefinitely. Although the numbers stated here cannot yet be
regarded as an accurate prediction, one can interpret the above
result such, that the number of fundamental particles is finite
and is expected to lie not too far outside range given in equation
(\ref{eq:f:Planck:2}). Thus we have a good chance to discover a
unified description of nature, encompassing all known forces and
matter states.

The lower value in the range given by equation
(\ref{eq:f:Planck:2}), $6270$, is quite close to the experimental
estimate of $f$, which can be obtained from equation
(\ref{eq:rho:fe}), using the temperature of the microwave
background radiation and the total matter density as input. With
$T_{CMBR} = 2.725 \, K$ and $\rho$ determined from the recent WMAP
data \cite{WMAP/cosmologicalParameters} we find the following
experimental estimate for the effective degrees of freedom at the
low energy scale:

\begin{equation}
f = \frac{60}{\pi^2} \frac{\rho}{T^4} \frac{\hbar^3}{(1+\frac{w_E+
\overline{w_E}}{2})} = \frac{18}{\pi^2} \frac{\rho}{T^4} \hbar^3
\approx 6366
\end{equation}

Therefore the assumption, that $r_0^2 = \beta \hbar$ is a nearly
universal area scale, only depending on the effective degrees of
freedom at a given energy scale via the relevant coupling
constants, has some observational justification.

\section{Does the holographic solution conserve the ratios of the energy-densities of the fundamental particle species?}

Quite interestingly $f/4$ of equation (\ref{eq:f:Planck:2}) is not
too far from the ratio of proton to electron mass ($m_p / m_e =
1836.15$). As the electron has four states (according to the
Dirac-equation), the ratio of proton to electron-mass is roughly
equal to the ratio of the {\em total} number of particle degrees of
freedom in the holostar-solution to the {\em four} degrees of freedom of
the electron.

At first sight this looks like quite a coincidence. However, there
is another curious coincidence in our universe today: The
energy-density of the microwave-background radiation is roughly
equal to the energy density of the electrons within a factor of
$2.5$ (if we assume that there is no dark matter). Both electrons
and photons are fundamental particles. If the near equivalence of
photon energy density to electron energy density is not just a
spurious feature of the universe in its present state\footnote{In
the standard cosmological model the energy-density of the photons
with respect to the electrons changes with time, as the photons
are "red-shifted" away and the ratio of photons to electrons is
expected to remain constant. However, in a non-homogeneous
universe with significant pressure it is not altogether clear, if
the number ratio of zero rest-mass particles to massive particles
in the cosmic fluid should remain constant. In fact, in the
holostar universe one can show (see \cite{petri/hol}) that this
ratio develops proportional to $\overline{n}_\gamma /
\overline{n}_e \propto m_e / T$ in the frame of the co-moving
observer, so that the red-shift of the photons is compensated by
their higher number-densities, keeping the ratio of the respective
energy densities constant.}, one wonders, whether the universe
might be constructed such, that the ratio of the energy-densities
of the different fundamental particle species should remain
approximately constant throughout its evolution, endowing every
fundamental degree of freedom with a well-defined energy-density.

In \cite{petri/hol} it was shown, that for the geodesic motion of
massless and massive particles, and for some particular cases of
non-geodesic motion of massive particles, the ratios of the
energy-densities of the particles are conserved in the interior
holostar space-time. For these particular cases the conjecture has
the status of being proved.

There are indications that such a conjecture is also valid with
respect to the distribution of electromagnetic (and rotational)
energy in the interior of a charged and/or rotating holostar: In
the model for a charged holostar discussed in \cite{petri/charge}
the ratio of electromagnetic energy density $\rho_{em} \propto 1/
r^2$ to the total energy density $\rho = 1 / (8 \pi r^2)$ is
constant throughout the whole holostar's interior and is related
to the dimensionless ratio of the holostar's exterior conserved
charge $Q^2$ and boundary area $A$ via $\rho_{em} / \rho_{tot} =
Q^2 / r_h^2 = 4 \pi Q^2 / A = const$.

If one elevates the above stated conjecture to the status of a
general principle, one is more or less forced to regard the
nucleon, which quite definitely is not a fundamental particle, as
some sort of "dump" for the frozen out fundamental degrees of
freedom at the Planck scale, which have become "locked away" in
the interior structure of the nucleon. According to the discussion
in section \ref{sec:asymmetry} at least  84 \% of the
energy-density in the ultra-relativistic particles at the
Planck-scale must be preserved in the subsequent evolution, due to
the profound matter-antimatter asymmetry at ultra-relativistic
energies. Using Occams razor (which allows us to ignore the
unsettled issues of cold dark matter or dark energy), where else
than into the nucleon could the energy density of these frozen out
degrees of freedom have gone to?

\subsection{An estimate for the ratio of the energy-density of photons to electrons}

If we take the conjecture seriously, that the universe preserves
the ratios of the energy-densities of its fundamental constituents
during its evolution, we should be able to estimate the ratio of
the energy-densities of electrons to photons by a thermodynamic
argument. When the temperature in the holographic universe reaches
the electron-mass threshold, we have $f_F = 5$ ultra-relativistic
fermionic particle species around: 3 flavors of neutrinos and two
helicity-states for the electrons. If right-handed neutrinos and
left-handed anti-neutrinos exist, we have to add three more
degrees of freedom, i.e. $f_F = 8$. The photons are counted as
$f_B = 1$, so $f_F / f_B \approx 5-8$. In this range, the relative
number-density of the electrons with respect to the photons is
given by $w_N \approx 2.6$, according to Table \ref{tab:thermo}.
The number-density of the positrons is roughly $\overline{w_N}
\approx 0.19$ that of the photons. The ratios of the
energy-densities of electrons {\em and} positrons to photons is
roughly $w_E + \overline{w_E} \approx 3.5$. After the annihilation
of the positrons roughly 86 $\%$ of the original energy-density
will "survive" in the left-over electrons. The $14 \%$ gone into
the annihilation is distributed among all left over particles,
i.e. neutrinos, photons and electrons.\footnote{In principle the
nucleons could also participate in the energy-transfer. As long as
the nucleons are much heavier than the electrons at the
electron-mass threshold, the energy-transfer to the nucleons will
be small.} The final result is, that the energy-density of the
left-over electrons with respect to the energy-density of the
photons should lie in the range

$$\frac{e_e}{e_\gamma} \approx 2.0 - 2.8$$

The lower value refers to the case, when the neutrinos are already
decoupled, so that the photons get the full share of the
annihilation energy ($14 \%$). When the neutrinos are not yet
decoupled, they will take a large fraction of the $14 \%$
annihilation energy, leading to the higher value.

Remarkably the observationally determined estimate of the ratio of the
energy-density of the electrons to the photons is quite close to
the above figure. When we estimate the electron contribution to
the total matter-density as determined by WMAP, under the
assumption that all of the matter is baryonic, and compare this to
the known energy-density of the CMBR, we get a ratio of $2.5$, as
was shown in \cite{petri/hol}.

\subsection{An estimate for the ratio of proton to electron mass}

With a similar argument one can estimate the proton to electron
mass ratio. We have seen that at ultra-high temperatures the total
number of particle degrees of freedom in the holostar-solution is
given by

$$f = 2^4 3^2 4 \pi \beta$$

At low temperatures the only fundamental  particle degrees of
freedom left are the electrons ($f_e =4$), the neutrinos ($f_\nu =
3 \cdot 2$) and the photons ($f_\gamma = 2$). The number of
"missing" degrees of freedom is given by

$$\Delta f = 2^4 3^2 4 \pi \beta - 12$$

In a spherically symmetric space-time mass-energy is conserved.
The mass-energy of the missing particle degrees of freedom must
show up somewhere. At low energies the natural candidate for the
missing degrees of freedom is the lightest surviving compound
particle, the proton. Therefore we can estimate the energy-density
of protons to electrons as follows

$$\frac{e_p}{e_e} \approx \frac{\Delta f}{4} = 2^2 3^2 4 \pi \beta - 3$$

However, this is the ratio of the energy-densities when the
electrons are still relativistic. When the electrons and positrons
finally annihilate, we have estimated in the previous section that
only roughly 84 \% of their original energy-density finds its way
to the surviving electrons, so that the ratios of the energy-densities after the annihilation of the positrons has to be corrected by this factor. At low temperatures the ratios of the energy-densities are nothing else than the ratios of the respective rest-masses, so that the ratio of proton to electron mass can be estimated as:

\begin{equation} \label{eq:mp/me}
\frac{m_p}{m_e} \approx \frac{2^2 3^2 4 \pi \beta - 3}{0.84}
\end{equation}

If we set $\beta$ to the value determined in \cite{petri/charge},
i.e. $\beta \simeq 4 \sqrt{3/4}$ we find:

\begin{equation}
\frac{m_p}{m_e} \approx 1862
\end{equation}

This is quite close to the actual value $m_p / m_e = 1836.15$.
Alternatively one could use equation (\ref{eq:mp/me}) to get
another experimental estimate for the fundamental area $r_0^2 =
\beta \hbar$.

In a more sophisticated treatment one would have to take into
account the bosonic and fermionic degrees of freedom separately.
Furthermore the neutron and the different chemical potentials of
neutrons and protons cannot be neglected.\footnote{Assuming that
the chemical potentials of $u$ and $d$ quarks are equal, and
assuming that the chemical potentials of the constituent quarks
can be summed up, the chemical potential of the neutron
$(\overline{d d} u)$ will be {\em opposite} to the chemical potential of
the proton $(u u \overline{d})$.}  However, with our limited
understanding of the holographic solution at the current time it
does not seem appropriate to stretch an order of magnitude
estimate far beyond its already limited range of credibility.

\section{Supersymmetry}

We have seen in the previous section, that the supersymmetric case
is special in the sense, that the ratio of the energy-densities of
fermions (including the anti-fermions) to bosons takes on an
integer value. The chemical potential per temperature of the
fermions in the supersymmetric case is given by

$$u_{ss} = \frac{\pi}{\sqrt{3}} \simeq 1.8138$$

For this value of $u$ we find:

\begin{equation} \label{eq:energy:ratio}
\frac{E_F + \overline{E_F} } {2 E_B} = \frac{Z_{F,3}(u_{ss}) +
Z_{F,3}(-u_{ss}) }{2 Z_{B,3}(0)} = \frac{w_E + \overline{w_E}}{2} =
\frac{7}{3}
\end{equation}

A similar relation holds for the entropy-densities:

\begin{equation} \label{eq:entropy:ratio}
\frac{S_F + \overline{S_F} } {2 S_B} = \frac{w_N \sigma_F +
\overline{w_N \sigma_F} } {2 \sigma_B} = \frac{3}{2 }
\end{equation}

With the above result one can relate $u_{ss}$ to the thermodynamic
parameters of the system:

\begin{equation} \label{eq:u:ss}
u_{ss} = \frac{20}{9} \frac{e_B}{\Delta n_F \, T} =
\frac{\pi^4}{27 \zeta(3)} \frac{n_\gamma}{\Delta n_F} \simeq
3.0013 \frac{n_\gamma}{\Delta n_F}
\end{equation}

$\Delta n_F$ is the fermion number density, i.e. the difference of
the number of fermions minus anti-fermions per unit volume, $e_B$
is the energy density of a single bosonic degree of freedom and
$n_\gamma$ is the number-density for a photon gas according to the
Planck distribution (g=2). We find that in the super-symmetric
phase the fermion number-density $\Delta n_F$ is roughly a factor
of $1.66$ higher than the boson number density $n_\gamma$.

Not quite unexpectedly equations (\ref{eq:energy:ratio},
\ref{eq:entropy:ratio}) guarantee, that the free energy comes out
zero, although the thermodynamic relations for an
ultra-relativistic photon gas imply that its free energy-density
is negative $f_B = e_B - s_B T = -e_B/3$, due to the well known
relation between energy- and entropy-density for a photon gas:

$$s_B T = \frac{4}{3} e_B$$

If the free energy shall be zero $s T / e = 1$ is required. We get
this by pairing any two bosonic degrees of freedom with a
fermionic particle anti-particle pair. According to equation
(\ref{eq:entropy:ratio}) the total entropy density is enhanced by
a factor $5/2$ by this pairing

$$s = s_B(1+\frac{3}{2})$$

whereas the total energy density is multiplied by a factor $10/3$

$$e = e_B (1+\frac{7}{3})$$

The relation between the total energy- and entropy-density can be
expressed in terms of the relation for a photon gas:

$$\frac{s T}{e} = \frac{ \frac{5}{2} s_B T}{ \frac{10}{3} e_B} = \frac{3}{4} \frac{s_B T}{e_B} = 1$$

\section{\label{sec:membrane}Thermodynamics of the membrane}

The simple model of an ultra-relativistic fermion and boson gas,
subject to the interior spherically symmetric metric $g_{rr} =
r/r_0$, reproduces the Hawking-entropy and -temperature, therefore
giving a microscopic statistical explanation for the origin of the
Hawking-entropy and -temperature, which fits well into the
theoretical framework that has been developed for black holes over
the last decades.

So far only the physics of the holostar's interior has been
discussed. At least from the viewpoint of an exterior observer the
properties of the membrane cannot be neglected. The surface
pressure of the holostar's membrane carries a stress-energy equal
to the holostar's gravitating mass. Furthermore, the membrane
might substantially contribute to the entropy.

From the point of view of string-theory the properties of the
membrane are quite easily explained. The interior strings are
attached to the boundary membrane. Each string segment occupies a
surface patch of of exactly one Planck area. The membrane has
similar properties to that of a 2D-brane in string-theory. It has
surface-pressure, but no interior mass-energy.

In this section I will try to interpret the properties of the
membrane in terms of particles. It might turn out, that this is
not the correct approach, and that the final answer has to be
sought purely in the context of string theory. Yet the particle
interpretation allows us to give some fairly self-consistent and
(apparently) sensible answer to the questions which have been
skipped in the previous sections, i.e. what contribution the
membrane can or will make to the gravitating mass and the entropy
of the holostar in the framework of the thermodynamic model that
has been developed in the previous sections. Keep in mind that
this section is somewhat speculative, as not much is known about
self-gravitating matter-states which are effectively confined to a
two-dimensional surface of spherical topology.

According to the holostar-equations the membrane has a large
surface pressure. This property would be difficult to explain, if
the membrane consisted only of weakly or non interacting
particles.\footnote{In fact, the zero energy-density in the
membrane suggests, that the membrane does not consist out of
particles at all, and that it is a pure string structure.} The
"forces" holding the membrane together must be strongly attractive
and presumably long-ranged. Therefore photons or fermions seem not
very well suited candidates in order to explain the properties of
the membrane. Presumably the membrane consists of a gas or fluid
of strongly interacting bosons.

A natural candidate for such a boson is the graviton. If the
number of bosons in the membrane is comparable to the number of
fermions inside the holostar\footnote{Supersymmetry suggests a
direct correspondence between fermions and bosons. If
supersymmetry is relevant for large holostars, the assumption that
the number of particles comprising the interior of the holostar
(presumably dominated by fermions) should be proportional to the
number of particles in the membrane (possibly dominated by bosons)
doesn't seem too far fetched.} the bosons will be very close to
each other. Their mean separation will be of the order of the
Planck-length. At such small distances gravity will have become a
very strong force. Therefore mutual interactions of gravitons at
close range could provide the "glue" holding the membrane
together. Note also, that gravitons, being spin-2 particles, have
the same transformation properties as the patches of a
two-dimensional surface.

The radial metric-coefficient $g_{rr}$ of the holostar space-time
attains its largest value at the position of the membrane. The
effective potential for the motion of massive and massless
particles has a global minimum at the position of the membrane.
Therefore the movement of particles in the radial direction is
extremely "inhibited", whereas movement in the tangential
direction, i.e. within the membrane itself, can be considered to
take place essentially unhindered. For large holostars, the local
movement of any particle in the vicinity of the membrane will be
effectively constrained to its two-dimensional surface area.

The membrane is the minimum of the effective potential of the
holostar. If we assume some weak, but non-zero interaction
(friction) between the particles in the vicinity of the membrane,
particles will collect in the membrane as the location of minimum
energy. Whereas bosons have no problem to occupy the same volume,
fermions are subject to the exclusion principle. A membrane which
consists of a large number of particles separated by a proper
distance of roughly Planck-size will quite likely contain a vast
number of bosons, but essentially no fermions.

The total number of particles of an ideal gas of bosons moving
freely in a two-dimensional surface of proper area $A$ is given by:

\begin{equation} \label{eq:NBose}
N = N_0 + \frac{f}{2 \pi \hbar^2} T^2 A Z_{B,1}(u_B)
\end{equation}

$N_0$ is the number of bosons condensed into the ground state, $f$
is the number of degrees of freedom of the bosons in the membrane
(for gravitons: 2), $T$ is the temperature at the membrane and
$Z_{B,1}$ is one of the integrals defined in section
\ref{sec:fermionBosonGas}.

$A \, T^2$ is proportional to $r_h$ and $Z_{B,1}(u_B)$ attains its
maximum value for $u_B = 0$. Under the assumption that the total
number of particles, $N$, in the membrane is proportional to
$r_h^2$, and disregarding the ground state occupation $N_0$, the
left side of equation (\ref{eq:NBose}) will grow much faster than
the right side. At the Bose-temperature the total number of
particles, $N$, will exceed the maximum value possible for the
second term on the right side of equation (\ref{eq:NBose}). At
this point the occupation of the ground state, $N_0$, must become
macroscopic.

The transition temperature $T_B$ from microscopic ground state
occupation ($N_0 \sim 0$) to macroscopic occupation can be
calculated from equation (\ref{eq:NBose}) by setting $N_0 = u_B =
0$. One finds:

\begin{equation} \label{eq:TBose}
T_B^2 = \frac{N}{A}\frac{2 \pi \hbar^2}{f Z_{B,1}(0)} = \frac{3
\hbar}{ f \pi \sigma} \approx 0.3 \frac{\hbar}{f}
\end{equation}

where $A = 4 \sigma N \hbar$ was used, assuming that the number of
bosons in the membrane is equal to the number of the holostar's
interior particles. The Bose-temperature of the two-dimensional
membrane is independent of $r_h$ and of order of the
Planck-temperature $T_{Pl} = \sqrt{\hbar}$. For reasonable values
of $f_B$ and $f_F$ the numerical factor in (\ref{eq:TBose}) will
lie in the interval from $0.285$ and $0.303$. With two degrees of
freedom for the gravitons one finds that the Bose temperature of
the membrane is roughly one third of the Planck temperature. The
mean energy of a boson constrained to a two-dimensional surface is
$E = (12 \zeta(3) / \pi^2)\, T \approx 4.4 T$. Therefore, at the
Bose-temperature the mean energy of the bosons in the membrane
slightly exceeds the Planck energy.

For large holostars the local temperature of the membrane will be
far less then its Bose-temperature, due to the
$1/\sqrt{r}$-dependence of the local temperature. Even for a
holostar of Planck-size Bose condensation of the membrane is
likely.\footnote{However, it will not be possible to regard the
membrane as a continuous surface, as has been done in the
semi-classical approach in this paper.} With the possible
exception of very small holostars, all of the bosons comprising
the membrane will be condensed into the ground state. Therefore
the membrane will not contribute to the entropy.

On the other hand, the membrane's contribution to the energy
cannot be neglected. The ground state energy for a single boson
will be somewhat larger than the energy of a standing wave with a
wavelength comparable to the proper circumference of the holostar:

\begin{equation} \label{eq:EgroundBose}
E_0 \simeq \frac{\hbar}{2 \pi r_h} \xi_0
\end{equation}

$\xi_0 \approx 3$ is the constant for the lowest vibrational mode
of the membrane. Its exact value can be determined by solving the
equations for a vibrating spherical membrane.

A rough estimate of the total energy of the membrane can be
attained by multiplying the ground state energy $E_0$ per boson
with the number of particles from equation (\ref{eq:Nclassic}).
Using relation (\ref{eq:s}) we get:

\begin{equation} \label{eq:EMembrane}
E_{m} = E_0 N \simeq \frac{r_h}{8 \pi^{3/2}}
\left(\frac{f}{\beta}\right)^{\frac{1}{4}} \xi_0 = M
\frac{\xi_0}{\sigma}
\end{equation}

$M$ is the gravitating mass of the holostar. The bosonic energy of
the membrane is comparable to the gravitating mass of the
holostar, giving further support to the holographic principle.
However, $E_m$ does not include the gravitational binding energy
of the bosons within the membrane. The classical holostar
equations predict a zero energy-density within the membrane, so
that one expects that the binding energy is exactly opposite to
$E_m$.

The argument can be turned around. From the holostar solution it
is known that the tangential pressure of the membrane has a
stress-energy "content" equal to the holostar's gravitating mass.
Under the assumption, that (i) the membrane consists of bosons,
(ii) each boson in  the membrane is in its ground state (with a
wavelength roughly equal to the proper circumference of the
holostar) and (iii) the mass energy within the membrane (excluding
gravitational binding energy) can be estimated by simply summing
up the individual boson energies, the total number of bosons
constituting the membrane can be estimated via equations
(\ref{eq:EgroundBose}) and (\ref{eq:EMembrane}):

\begin{equation} \label{eq:NMembrane}
N_{m} \simeq M / E_0 \simeq \frac{\pi r_h^2}{\hbar}
\frac{1}{\xi_0}= \frac{\sigma}{\xi_0} N
\end{equation}

This estimate gives the same order of magnitude for the number of
bosons in the membrane, $N_m$, and the number of particles within
the holostar's interior, $N$.

\section{\label{sec:T:zero}The zero temperature case}

In this section I discuss a "zero-temperature" holostar. I have
put in this section rather for the completeness of coverage than
being convinced that a zero-temperature holostar exists. However,
the reader may judge for himself.

In the zero-temperature case the bosonic contribution to the
(interior) mass-energy density and entropy can be neglected with
respect to the fermions. At $T=0$ the holostar's interior should
be essentially free of bosons, all of which will have assembled in
the membrane as the state of lowest energy. This will not be an
option for the fermions, which are subject to the exclusion
principle.

In the zero-temperature case the fermi-distribution in momentum
space is given by a Heavyside step-function, which is unity for
low momenta and falls off to zero abruptly at the fermi-momentum
$p_F$. The fermi-momentum is nothing else than the chemical
potential at $T=0$.

The number of relativistic fermions in an interior spherical shell
of volume $\delta V$ can be calculated as follows:

\begin{equation} \label{eq:dN1_T0}
\delta N = \frac{f}{2 \pi^2 \hbar^3} \delta V \frac{p_F^3}{3}
\end{equation}

where $p_F$ is the Fermi-momentum.

The energy in the shell is given by:

\begin{equation} \label{eq:E_T0}
\delta E = \frac{f}{2 \pi^2 \hbar^3} \delta V \frac{p_F^4}{4} =
\frac{3}{4} p_F \delta N
\end{equation}

The mean energy of a highly relativistic fermion within the shell
is lower than its fermi-momentum $p_F$, because all momenta up to
$p_F$ are occupied. In 3D-space the average momentum is $3/4 \,
p_F$.

The energy of the shell per proper volume must be equal to the
mass-energy density of the holostar solution. Therefore:

\begin{equation} \label{eq:Rho_T0}
\frac{\delta E}{\delta V} = \frac{f}{2 \pi^2 \hbar^3}
\frac{p_F^4}{4} = \frac{1}{8 \pi r^2}
\end{equation}

From this the fermi-energy $p_F$ can be read off:

\begin{equation} \label{eq:EF_T0}
{p_F}^4 = \frac{\pi \hbar^3}{f} \frac{1}{r^2}
\end{equation}

Inserting $p_F(r)$ from above into equation (\ref{eq:dN1_T0}) and
using equation (\ref{eq:dV}) for the volume element we can
determine the number of fermions within the shell:

\begin{equation} \label{eq:dN_T0}
\delta N = \frac{2}{3} \left(\frac{f}{\pi
\beta}\right)^\frac{1}{4} \frac{r \delta r}{\hbar} = \frac{1}{3
\pi} \left(\frac{f}{\pi \beta}\right)^\frac{1}{4} \delta S_{BH}
\end{equation}

$S_{BH}$ is the Bekenstein-Hawking entropy attributed to the shell.

According to the derivation in the last section we should now
compare the thermodynamic entropy with the Hawking entropy in
order to determine $\beta$. But there is a difficulty with this
approach: The thermodynamic entropy of a zero-temperature holostar
is zero. All states within the fermi-sphere are occupied. There is
just one microscopic configuration for such a degenerate
macroscopic state.

What seems to be possible, though, is to compare the momentum of
the fermions at the holostar's surface with the Hawking
temperature (at infinity). In order to do this, we have to
establish a relation between the local "temperature" and the
fermi-energy at the holostar's surface. Let us consider the
process, where a thin shell of particles is added to the holostar.
Any fermion in the newly added shell has an energy given by
equation (\ref{eq:E_T0}). The total energy of the shell is given
by:

\begin{equation}
\delta E = \frac{3}{4} p_F \delta N
\end{equation}

If any fermion carries an "intrinsic" entropy $\sigma_0$, the
entropy of the newly added shell will be given by

\begin{equation}
\delta S = \sigma_0 \delta N
\end{equation}

Combining this with the known thermodynamic relation $\delta S =
\delta E/T$, we find:

\begin{equation} \label{eq:pF:T=0}
p_F = \frac{4}{3} \sigma_0 T_0
\end{equation}

We can determine $\sigma_0$ by comparing the "temperature" $T_0$
at the holostar's surface with the Hawking temperature (both
temperatures compared at infinity):

\begin{equation} \label{eq:intrinsic:entropy}
\sigma_0 = 3 \pi \left( \frac{\pi \beta}{f}\right)^\frac{1}{4}
\end{equation}

I haven't found a way to give accurate numerical figures for
$\sigma_0$ in the zero temperature case, as was possible in the
case of non-zero temperature. This would require us to know both
$f$ and $\beta$. Furthermore, one runs into severe problems, when
one tries to interpret the "intrinsic" entropy $\sigma_0$ and the
"temperature" $T_0$ (derived from the fermi momentum) in a
thermodynamic sense. For instance, with $\mu_F = p_F$ the chemical
potential per "temperature" naively is given by

\begin{equation} \label{eq:u:intrinsic:entropy}
u_F = \frac{p_F}{T_0} = \frac{4}{3} \sigma_0
\end{equation}

One could now try to use the above relation to calculate the
intrinsic entropy $\sigma_0$ from the thermodynamic equations of
an ultra-relativistic gas of fermions at {\em non-zero}
temperature, assuming that the "intrinsic" entropy per fermion is
equal to the thermodynamic entropy given by equation
(\ref{eq:FS:FE}). This would allow us to express $\sigma_0$ as a
function of $u_F$. But this approach fails. The implicit equation
for $u_F$ has no solution. Furthermore, equation
(\ref{eq:u:intrinsic:entropy}) is not even symmetric in $u_F$, so
instead of getting two solutions $u_F$ and $-u_F$, which can be
interpreted as particle/anti-particle pair, we just get a
nonsensical imaginary result. The failure of this approach is not
quite unexpected. It doesn't really make sense to use the $T=0$
Fermi-distribution for the calculation of energy- and
number-densities and then set $T = p_F / u_F \neq 0$ in order to
get rid of the undesired result $S=0$ for the thermodynamic
entropy. Quite obviously it requires a considerable amount of
"creative cheating" in order to make a zero-temperature, zero
entropy holostar compatible with the Hawking entropy and
temperature relations.

An approach which quite likely is not correct either, but which at
least leads to some sensible numerical figures, is to compare the
entropy per fermion $\sigma$ in the simple model of section
\ref{sec:fermion:simple} (a holostar, whose interior consists only
out of fermions) with the "intrinsic entropy" $\sigma_0$ of a
particle in the "zero-temperature" holostar, which is given by
equation (\ref{eq:intrinsic:entropy}). The value of $\sigma$ is
given by equation (\ref{eq:s}). The ratio of both quantities turns
out as:

\begin{equation}
\frac{\sigma_0}{\sigma}=\frac{3}{4 \pi^\frac{1}{4}} \simeq 0.563
\end{equation}

Knowing $\sigma$ (for the non-zero temperature case), one might
then be able to determine $\sigma_0$ via the above ratio. In the
more sophisticated model of section \ref{sec:fermionBosonGas} the
entropy $\sigma$ for a (non-zero) temperature holostar consisting
only out of fermions has been shown to be $\sigma \simeq 3.38$, so
that $\sigma_0$ might be estimated as

$$\sigma_0 \simeq 1.90$$

With $\sigma_0$ the total number of particles $N$ can be
calculated. It turns out larger than in the non-zero temperature
case by roughly a factor of $1.8$ (assuming $\sigma \simeq 3.38$
for the non-zero temperature case):

\begin{equation} \label{eq:N:T0}
N = \frac{A}{4 \sigma_0 \hbar} \approx 0.13 \frac{A}{\hbar}
\end{equation}

Up to somewhat different constant factors the zero temperature
model produces essentially the same results as the non-zero
temperature model discussed in the previous sections. However, for
$T=0$ the thermodynamic entropy of the interior fermions is zero.
The membrane doesn't contribute to the entropy anyway. Therefore a
zero temperature holostar should have no appreciable thermodynamic
entropy, which is in gross contradiction to the Hawking
entropy-area law. Giving the fermions an "intrinsic" entropy can
solve the problem, but not in a truly satisfactory way. Therefore
I rate it doubtable that the the zero-temperature case is a
physically acceptable description for a compact self gravitating
object.

\section{Rotation}

In order to study collision or accretion processes involving the
new type holostar solutions it will be necessary to find a
solution that describes a rotating object.

Some properties of a yet to be found rotating axially-symmetric
holostar solution might be inferred from the spherically symmetric
solution. For a first approximation one could assume that the
holostar rotates stiffly, at least for small rotation speeds.
Unfortunately this requires infinitesimally small rotation rates
$d\varphi/dt$ for a large holostar, in order that the holostar's
surface doesn't rotate faster than the local velocity of light.

On the other hand, the event horizon of a Kerr black hole is known
to rotate stiffly, irrespective of the rotation speed. Although
there is no theorem like Birkhoff's theorem for axially-symmetric
space-times, it is not unreasonable to assume that the exterior
space-time of a rotating holostar is similar, if not identical, to
the Kerr-metric. If this is true, at least the surface of a
holostar should rotate stiffly. Furthermore one can expect that a
rotating holostar will have no differential rotation within any
interior spherical shell, although spherical shells with different
radial coordinate positions\footnote{Keep in mind that for a
rotating holostar it is not trivial to determine the geometric
interpretation of a chosen "radial coordinate".} might rotate
differentially with respect to each other. If there is
differential interior rotation, it is quite probable that an
interior observer will not be able to discern any peculiar angular
motion due to differential rotation of the interior shells. The
proper radial distance between adjacent spherical shells is huge,
due to the $r/r_0$-dependence of the radial metric coefficient.
Therefore, as long as the differential rotation doesn't depend
exponentially on the radial coordinate value, an interior observer
will not be able to notice the differential rotation of the shell
with respect to an adjacent shell, even if the proper distance to
the adjacent shell is equal to the Hubble-radius of the interior
observer. This is due to the fact, that the radial coordinate
distance between the shells becomes infinitesimally small for any
fixed proper radial distance.

It is quite clear, that a holostar's rotation rate is limited. It
must rotate less than the maximum rotation rate of a Kerr black
hole: A maximally rotating (extreme) Kerr black hole is
characterized by the property, that the proper velocity of the
(stiffly rotating) event horizon equals the velocity of light. For
the holostar solution this would imply that the rotation speed of
the membrane, which lies slightly outside the gravitational
radius, would exceed the velocity of light.

\subsection{A bound for the mean spin-alignment of the interior particles}

This upper bound enables us to get some valuable information with
respect to the interior structure of a rotating holostar. For this
purpose it is instructive to reflect on how the rotation will
manifest itself locally in the holostar's interior. The overall
rotation is expected to force the interior particles to align
their spins and orbital momenta along the rotation axis. I
assume, that the rotation doesn't change the number of interior
particles.\footnote{This requires that energy must be transferred
adiabatically to the holostar in order to spin it up from zero to
maximum angular momentum.} In this case the entropy (and surface
area) should remain constant. The maximum rotation will be
achieved, when the spins and orbital momenta of all interior
particles are aligned. The dominant particle species within a
holostar, at least with respect to the number of particles, are
ultra-relativistic particles. If we neglect the orbital momenta,
we can determine the fraction of the total angular momentum of a
holostar, $J_H$, due to the alignment of its spins:

\begin{equation} \label{eq:J:hol}
J_{H} = \overline{j} \hbar N = \frac{\overline{j}}{4 \sigma} A
\end{equation}

$\overline{j}$ is the expectation value of the spin quantum number
of the ultra-relativistic particles in direction of the exterior
rotation axis. For a maximally aligned holostar (meaning the $j_z$
component of the spins of all particles point into the direction
of the exterior rotation axis) $\overline{j}$ will be equal to the
sum of the spin-quantum numbers of all the particles composing the
holostar divided by the total number of particles. For a large
holostar approaching the size of the universe, the dominant
interior (relativistic) particle species will be the neutrinos
with $j_F = 1/2$ and possibly photons with $j_B = 1$.

If orbital angular momentum is included, the total angular
momentum of the holostar will be higher than the value in equation
(\ref{eq:J:hol}). Alternatively $\overline{j}$ can be interpreted
as the expectation value of the spin of the particles in direction
of the rotation axis, including the mean orbital momentum of the
particles (if there is any). To get an estimate of the maximum
possible rotation rate of a holostar we can compare the angular
momentum of a maximally aligned holostar to that of a maximally
rotating (extreme) Kerr black hole.

For an extreme Kerr-Black hole there is a definite relationship
between its angular momentum, $J_K$, and surface area:

$$J_{K} = \frac{A}{8 \pi}$$

The holostar's angular momentum can never exceed the angular
momentum of an extreme Kerr-black hole, which requires $J_H <
J_K$. Comparing the two angular momenta we find:\footnote{Note
that both objects are compared assuming that their surface areas
are equal. This seems the most natural assumption. A rotating
holostar, whose number of particles is equal to the non-rotating
case, should have the same entropy and thus the same surface area.
We therefore should compare the rotating holostar with a Kerr
black hole of the same area.}

$$\frac{J_{H}}{J_{K}} = 2 \overline{j} \frac{\pi}{\sigma} < 1$$

This can be interpreted as a bound on the mean spin quantum
number of the interior particles of a rotating holostar:

\begin{equation} \label{eq:j:bound}
\overline{j}  < \frac{1}{2} \frac{\sigma}{\pi} \approx \frac{1}{2}
\cdot \left( 1.004 \ldots 1.07 \right)
\end{equation}

In our simple thermodynamic model the value of $\sigma$, the
entropy per fermion, is very close to $\pi$. The exact value of
$\sigma$ depends on the ratio between the bosonic and the
fermionic degrees of freedom and on how the chemical potentials of
bosons and fermions are related. For all reasonable values of
$f_F$ and $f_B$, the ratio $\sigma / \pi$ is larger than $1$
although only by a small percentage, as can be seen from the
tables in section \ref{sec:Tables}.

For the realistic thermodynamic model, i.e. when the chemical
potential of the bosons is zero (Table \ref{tab:thermo}) and for
realistic values of $f_B / f_F < 3000$, $\sigma / \pi$ attains its
maximum for $f_B/f_F = 0$ and its minimum at roughly $f_B / f_F =
8$:

\begin{tabbing}
$(\sigma /\pi)_{max}$  \= $\approx 1.076$ \quad \= for $f_B = 0$ \\
$(\sigma / \pi)_{min}$  \> $\approx 1.004$ \> for $f_B /f_F
\approx 8$
\end{tabbing}

For the "abnormal supersymmetric phase", i.e. when the chemical
potential of the bosons is opposite in sign to the chemical
potential of the fermions (Table \ref{tab:super}), we find:

\begin{tabbing}
$(\sigma / \pi)_{min}$ \= $\approx 1.067$ \quad \= for $f_B = 0$ \\
$(\sigma / \pi)$ \> $\approx 1.073$ \quad \> for $f_B = f_F$
\end{tabbing}

In any case, whatever the combination of $f_F$ and $f_B$ and the
relation between the chemical potentials might be, one can come
very close to the angular momentum of a maximally rotating
Kerr-Black hole, by simply aligning the spins of the
ultra-relativistic fermions (and bosons) of the spherically
symmetric holostar solution. If the (mean effective) spin quantum
number of the ultra-relativistic particles within the holostar is
larger than $1/2$, it is possible to exceed the angular momentum
of a maximally rotating Kerr-black hole, simply by aligning a
large proportion of the spins, as can be seen from equation
(\ref{eq:j:bound}). If the interior particles have an appreciable
orbital angular momentum, which was neglected in the determination
of $J_{H}$, the Kerr-limit will be exceeded, even if all of the
ultra-relativistic particles are spin-1/2 particles (i.e.
particles with the lowest non-zero spin quantum possible). This is
physically unacceptable. The outer regions of the holostar would
rotate with a velocity larger than the speed of light.

\subsubsection{interior particle spin $j=1/2$}

Under the assumption, that the holostar consists only out of spin
1/2 fermions, aligning the spins of all particles already yields a
total angular momentum that is quite close to the maximum angular
momentum possible, $J_K$:

\begin{equation} \label{eq:J:max}
J_{H} = \frac{\pi}{\sigma} J_{K} = 0.9297 J_{K}
\end{equation}

$J \approx 0.9 \, J_K$ is quite close to the rotation rates
expected for black holes formed from realistic gravitational
collapse or accretion processes, when the angular momentum of the
collapsing matter is taken into account. In fact, very recently
the mass and angular momentum of the black hole in the center of
our galaxy has been measured by analyzing the frequency spectrum
of X-ray flares coming from the galactic center
\cite{Aschenbach/2004}. The measurements allow a very precise
determination of the angular momentum variable $a = J / J_{K}$
($a$ denotes the ratio of the angular momentum of a black hole to
its maximum possible value, which is given by the angular momentum
of an extreme Kerr black hole). The data allow two solutions,
characterized by a low ($M = 2.79(4) \cdot 10^6 M_\odot$) and a
high ($M = 4.75(7) \cdot 10^6 M_\odot$) gravitational mass of the
black hole. Although the masses differ by a factor of almost two,
the angular momentum parameters are nearly identical, $a =
0.9937(7)$ for the low and $a = 0.991(2)$ for the high mass
solution. Both values are very close to the prediction $a =
0.9297$ for a maximally aligned fermionic holostar (Eq.
(\ref{eq:J:max})).

The argument can be turned around. There appears to be no basic
physical law that forbids the interior particles of a holostar to
align their spins along a common axis. If this is so, the interior
(massless) particles of a rotating holostar should be mostly spin
$1/2$ particles or a mixture of spin $0$, spin $1/2$ and spin $1$
particles, with not too high a contribution of particles with
higher spin, otherwise the Kerr-limit would be exceeded.\footnote{
This is quite similar to the situation in loop quantum gravity,
where the number of states of the area operator (within a given
small area range) is dominated by spin 1/2 links for large areas.}
Furthermore, the interior particles of a rapidly rotating holostar
cannot have high orbital angular momenta, otherwise the Kerr-limit
would be exceeded even if the holostar consisted exclusively out
of spin-1/2 particles.

\subsubsection{interior particle spin $j=1/2$ and $j=1$}

If the holostar contains one massless fermionic and one massless
bosonic species, such as the neutrino and the photon, it not
possible to align all particles. Let us construct a very simple
example. Consider the case $f_F = f_B$. Let all of the fermions
have spin $\overline{j_F} = 1/2$ and all of the bosons spin
$\overline{j_B} = 1$. According to the formula and tables given in
section \ref{sec:Tables}, the number density of the fermions +
anti-fermions per proper volume is a factor of $w = (w_N +
\overline{w_N})/2 \simeq 1.7877$ higher than the number density of
the bosons.

If the spins of all particles are aligned, the mean expectation
value of all spins in the direction of the alignment-axis is given
by:

\begin{equation} \label{eq:j:super}
\overline{j} = \frac{w}{1+w} \overline{j_F} + \frac{1}{1+w}
\overline{j_B} = 0.6794
\end{equation}

This is larger than the bound of equation (\ref{eq:j:bound}),
evaluated for $\sigma = 3.2299$:

\begin{equation} \label{eq:jmax:super}
\overline{j} < \frac{1}{2}\frac{\sigma}{\pi} = 0.5141
\end{equation}

This bound cannot be exceeded, therefore either it is impossible
to align all of the spins or there must be a significant fraction
of spin-0 bosons. If only the fermions are aligned, we get:

\begin{equation}
\overline{j} = \frac{w}{1+w} \frac{1}{2} = 0.3206
\end{equation}

so that angular momentum of the holostar due to the alignment of
all of its spin-1/2 particles is roughly 62\% of the maximum spin.

It seems awkward to postulate, that the fermionic spins can be
aligned along a common axis and the bosonic spins not. If one
aligns all spins, there is no problem when the holostar consists
only out of fermions, whereas for equal fermion and boson numbers
the spin-limit is exceeded (with the assumption that all fermions
have spin-1/2 and all bosons spin-1). If one knows how the spins
are distributed among the different particle-species, such as in
certain supersymmetric models, it is possible to check whether
aligning all of the spins violates the spin-limit $j_{max} =
\sigma / (2 \pi)$. With the simplified assumption that all bosons
have spin-1 and all fermions spin-1/2, we find that there must be
at least 10.45 fermionic degrees of freedom per bosonic degree of
freedom, in order that the spin-limit is not exceeded.

\subsection{supersymmetric models}

An interesting value is the spin-limit for equal bosonic and
fermionic degrees of freedom. According to equation
(\ref{eq:jmax:super}) its value is given by $j_{ss} = 0.5141$. One
might use this value in order to restrict the various
supersymmetric models. For example, $N=8$ supersymmetry has a
mean-spin of the fermions of $\overline{j_F} = 5/8$ and a mean
spin of the bosons $\overline{j_B} = 15/32$. Using equation
(\ref{eq:j:super}) and replacing $\overline{j_F} \rightarrow 5/8$
and $\overline{j_B} \rightarrow 15/32$ leads to $\overline{j} =
0.569$, which exceeds the limit, whereas $N=4$ supersymmetry leads
to $\overline{j} = 0.410$, which is within the limit.

In any case, the spin-limit provides quite a stringent constraint,
which can be violated quite easily, even without taking the
angular momentum of the particle's motion into account. Therefore
we are led to the assumption, that a rotating holostar most likely
acquires its angular momentum predominantly due to the alignment
of its interior particles. There should be neither an appreciable
contribution from the higher spin particles, such as spin-1
bosons, nor a significant angular momentum contribution from the
interior particles. Note also, that if the angular momentum value
of a large charged holostar can be interpreted as the alignment of
its charged spin 1/2 particles along a common axis, this might be
an "explanation" for the $g$-factor of two for a rotating charged
Kerr-Newman black hole.

\subsection{the zero temperature case}

How does a zero-temperature holostar, as discussed in section
\ref{sec:T:zero} fit into this picture? For a zero temperature
holostar one finds a numerically different result for the maximum
spin expectation value:

$$\frac{{J_H}}{{J_K}} = 2 \overline{j} \frac{\pi}{\sigma_0} \simeq 3.32 \overline{j} < 1$$

or

$$\overline{j} \approx < \frac{1}{3}$$

If more than roughly two thirds of the spin $1/2$ fermions are
aligned along a particular axis, the angular momentum of the
zero-temperature holostar will exceed the angular momentum of the
extreme Kerr-solution. If the interior particles have higher spin
or orbital angular momentum, such a situation will occur at even
lower relative alignments. Rotation rates exceeding the extreme
Kerr-rate are physically not acceptable. On the other hand, from
the viewpoint of the interior observer there seems to be no good
reason, why not all of the spin's can be aligned along a common
axis. Therefore the above finding might be interpreted as further
evidence, that a zero-temperature black hole is not physically
realized.

\subsection{local CP-violation}

Quite interestingly, a rotating holostar could be a natural cause
for an (apparent) C and/or P violation. As has been shown in
\cite{petri/hol} the motion of any "ordinary" particle in the
holostar becomes nearly radial and outward directed, whenever the
particles can be considered as non-interacting. Such a
non-interaction condition is evidently satisfied by the neutrinos
in the outer regions of a holostar of the size of the universe. If
the spins of the radially outward moving neutrinos are forced to
line up with respect to a given exterior rotation axis, the
holostar will be divided into two half-spheres with a distinct
matter-antimatter asymmetry (if the neutrinos have just one
definite helicity state as assumed by the Standard Model). Under
the alignment-constraint imposed by the exterior rotation axis the
antineutrinos, with spin in direction of flight, will
preferentially move in the direction of the external rotation
axis, whereas the neutrinos with opposite helicity will
preferentially move in the opposite direction. If the holostar
attains its maximum rotation rate, i.e. all of the fermion spins
are aligned with respect to the rotation axis, there will be an
almost perfect neutrino/antineutrino asymmetry between the two
half-spheres. Neutrinos moving radially outward will dominate one
half-sphere, anti-neutrinos the other.

\section{Discussion and Outlook}

A simple thermodynamic model for the holostar solution has been
presented which fits well into the established theory of black
holes. From the viewpoint of an exterior observer the holostar
appears very similar to a classical black hole. The modifications
are minor and only "visible" at close distance: The event horizon
is replaced by a two dimensional membrane with high tangential
pressure, situated roughly a two Planck coordinate lengths outside
of the gravitational radius. The pressure of the membrane is equal
to the pressure derived from the so called "membrane paradigm" for
black holes \cite{Thorne/mem}, which guarantees, that the
holostar's action on the exterior space-time is indistinguishable
from that of a same-sized black hole. The membrane has zero
energy-density, as expected from string theory. The interior
matter is singularity free and can be interpreted as a radial
collection of classical strings, attached to the holostar's
spherical boundary membrane. Each string segment attached to the
membrane occupies a membrane segment of exactly one Planck area.
The string tension falls with radius and is inverse proportional
to the string length, as measured by an asymptotic observer at
spatial infinity.

The interior matter state can be interpreted in terms of
particles. In this paper a very simple thermodynamic model of an
ideal ultra-relativistic gas was discussed. Temperature and
entropy of the holostar are of microscopic origin and exactly
proportional to the Hawking temperature and -entropy. The number
of interior particles within the holostar is proportional to the
proper area of the boundary-membrane, measured in Planck units,
indicating that the holographic principle is valid for compact
self-gravitating objects of arbitrary size.

The surface temperature of a holostar measured at infinity is
proportional to the Hawking temperature. By this correspondence
one can set up a specific relation between the Hawking temperature
(measured at infinity), the interior radiation temperature and the
interior matter density. This correspondence allows an
experimental verification of the Hawking-temperature law from the
holostar's interior. By comparing the CMBR-temperature to the
total matter density of the universe the Hawking temperature law
has been experimentally verified to an accuracy of roughly 1 \%.
However, the numerical verification depends on the formula given
in \cite{petri/charge} for $\beta$ (Eq. \ref{eq:beta:running}),
which stills lacks a formal derivation.

The holostar solution has no singularity and no event horizon.
Information is not lost: The total information content of the
space-time is encoded in its constituent matter, which can consist
out of strings or particles. Unitary evolution of particles is
possible throughout the full space-time manifold. Every
ultra-relativistic particle carries a definite entropy, which can
be calculated when the number of ultra-relativistic fermionic and
bosonic degrees of freedom is known.

The thermodynamics of the holostar solution indicate, that
chemical potentials play an important role in the stabilization of
self gravitating systems. The holostar requires a non-zero
chemical potential, proportional to the local radiation
temperature, of at least one fermionic ultra-relativistic species.
Different thermodynamic models can be constructed, which are
characterized by the ratio of bosonic to fermionic degrees of
freedom.

The non-zero chemical potential of the ultra-relativistic fermions
acts as a natural cause for a significant matter-antimatter
asymmetry in a spherically symmetric curved space-time. For any
ultra-relativistic fermionic particle there are never more than
0.09 anti-particles, meaning the total fraction of anti-fermions
is always less than 8 \% of the total number of particles and
anti-particles of a given species. Therefore at least 84 \% of the
original energy-density of an ultra-relativistic fermion gas
"survives" in ordinary fermionic matter, if the temperature falls
below the mass-threshold and less than 8 \% anti-fermions
annihilate with their matter counterparts.

Although the interior structure of the non-rotating holostar can
be considered as fairly well understood, not much is known about
the membrane and the astrophysically interesting case of a
rotating holostar. These two topics present themselves as a very
interesting themes of future research. Some arguments with respect
to the interior structure of the membrane and the properties of a
- yet to be found - rotating holostar solution were given. It has
been proposed, that the membrane consists of a gas of bosons at a
temperature far below the Bose-temperature of the membrane,
forming a single macroscopic quantum state. Evidence has been
presented that a rotating holostar will acquire its angular
momentum preferentially by the alignment of its interior particles
along the rotation axis. Whether these tentative "predictions"
point into the right direction can only be answered by future
research.

Having two or more solutions for the field equations (black hole
vs. holostar) makes the question of how these solutions can be
distinguished from each other experimentally an imminently
important question. Can we find out by experiment or observation,
which of the known solutions, if any, is realized in nature? At
the present time the best argument in favor of the holostar
solution appears to be the accurate measurement of the Hawking
temperature via the CMBR-temperature and the matter-density of the
universe.

Yet it would be helpful if more direct experimental evidence were
available. Due to Birkhoff's theorem the holostar cannot be
distinguished from a Schwarzschild black hole by measurements of
its exterior gravitational field. But whenever holostars come
close to each other or collide, their characteristic interior
structure should produce observable effects, which deviate from
the collisions of black holes. Presumably a collision of two
holostars will be accompanied by an intense exchange of particles,
with the possible production of particle jets along the angular
momentum axis.

In accretion processes the membrane might produce a noticeable
effect. The rather stiff membrane with its high surface pressure
might be a better "reflector" for the incoming particles, than the
vacuum-region of the event horizon of a Schwarzschild-type black
hole. There are observations of burst-like emissions from compact
objects, which are assumed to be black holes because of their high
mass ($M > 3-5 M_{\odot}$), but that have spectra rather
characteristic for neutron stars, i.e. of particles "hitting" a
stiff surface. A more accurate observation of these objects might
provide important experimental clues to decide the issue.

A high density of bosons in the membrane, with a mean separation
comparable to the Planck-length, might induce copious
particle-interchange reactions, similar to what is expected when
particles cross a so called domain-wall. For holostars of
sub-stellar size ($r_h \approx 1 \, km)$ the local temperature at
the membrane becomes comparable to the nucleon rest mass energy. A
rather hot particle gas at the position of the membrane could
produce noticeable effects with respect to the relative abundances
of the "reflected" particles, due to high energy interactions with
the constituent particles of the membrane or the holostar's
interior.

On the other hand, the extreme surface red-shifts on the order of
$z \approx 10^{20}$ for a solar mass holostar, and larger yet for
higher mass objects ($z \propto \sqrt{M}$), might not allow a
conclusive interpretation of the experimental data with regard to
the true nature of any such black hole type object.

The most promising route therefore appears to be, to study the
holostar from its interior. In \cite{petri/hol} it has been
demonstrated, that the holostar has the potential to serve as an
alternative model for the universe. The recent WMAP-measurements
have determined the product of the Hubble constant $H$ times the
age of the universe $\tau$ to be $H \, \tau \simeq 1.02$
experimentally with $H = 71 \, (km/s)/Mpc$ and $\tau = 13.7 \,
Gy$. The holostar solution predicts $H \, \tau = 1$ exactly. There
are other predictions which fit astoundingly well with the
observational data. This in itself is remarkable, because the
holostar-solution has practically no free parameters. It's unique
properties arise from a delicate cancelation of terms in the
Einstein field equations, which only occurs for the "special"
matter density $\rho = 1 / (8 \pi r^2)$, leading to the "special"
radial metric coefficient $g_{rr} = r/r_0$ (see \cite{petri/bh}
for the derivation). That the holostar solution with its
completely "rigid" structure has so much in common with the
universe as we see it today, either is the greatest coincidence
imaginable, or not a coincidence at all.

With the holostar solution we have a beautifully simple model for
a singularity free compact self gravitating object, which is
easily falsifiable. Its metric and fields are simple, its
properties are not. It is an elegant solution, as anyone studying
its properties will soon come to realize. However, in science it
is experiments and observations, not aesthetics, that will have to
decide, which solution of the field equations has been chosen by
nature. It is our task, to find out. The work has just begun.


\end{document}